\documentclass[12pt]{article}
\usepackage{jheppub} 
\usepackage{mathrsfs,rotating,amsmath,amsthm,amsfonts,mathtools,booktabs,amssymb,wasysym,caption}
\usepackage{hyperref}
\usepackage{natbib}
\usepackage[table,xcdraw,dvipsnames]{xcolor}
\usepackage{graphicx}
\usepackage[utf8x]{inputenc}
\usepackage[english]{babel}
\usepackage{appendix}
\usepackage{libertine}
\usepackage{lmodern}
\usepackage{comment}

\def\be{\begin{equation}}
\def\ee{\end{equation}}
\def\ba{\begin{align}}
\def\ea{\end{align}}	
\newcommand{\norm}[1]{{|| #1 \, ||}}
\newcommand{\sfrac}[2]{{\textstyle\frac{#1}{#2}}}
\renewcommand{\Re}{{\rm Re}}
\renewcommand{\Im}{{\rm Im}}

\newcommand{\C}{\tilde G_c}

\title{Microcausality without Lorentz invariance} 

\author[a]{Lam Hui,}
\author[a]{Alberto Nicolis,}
\author[b]{Alessandro Podo,}
\author[a]{and Shengjia Zhou}

\affiliation[a]{Physics Department and Center for Theoretical Physics,\\
  Columbia University, New York, NY 10027, USA}
  
\affiliation[b]{Institut des Hautes \'Etudes Scientifiques, 91440 Bures-sur-Yvette, France}

\emailAdd{lh399@columbia.edu}
\emailAdd{a.nicolis@columbia.edu}
\emailAdd{podo@ihes.fr}
\emailAdd{sz2807@columbia.edu}

\abstract{Microcausality---the vanishing of commutators outside the lightcone---is a fundamental property of relativistic quantum field theories. We derive its implications for two-point functions of scalar operators on {\it Lorentz-breaking} states. We restrict to spatially homogeneous and isotropic states, at zero and finite temperature, such as finite-density states of matter and primordial inflationary states. In a mixed $(t, \vec k \, )$ representation, we find certain analyticity and exponential boundedness conditions, which we verify in a variety of examples. Crucially, we discuss how our conditions can be tested within the regime of validity of Lorentz-breaking low-energy effective field theories, clarifying the role of the group velocity of low-energy excitations. In the cosmological case, we derive a positivity condition on an EFT coefficient in an inflationary background.
Lastly, we comment on how microcausality can be used to constrain higher-point correlation functions, via suitable nested commutators.
}

\begin{document}
\maketitle

\section{Introduction}

One of the defining properties of relativistic quantum field theories (QFT's) is the fact that local observables commute at spacelike separations \cite{Strocchi},
\be
[{\cal O}_1(x), {\cal O}_2(y)] = 0 \qquad \mbox{if} \; (x-y)^2 < 0  \; ,
\ee
where ${\cal O}_1(x)$ and ${\cal O}_2(x)$ are any two bosonic, gauge-invariant local operators, and we are using the mostly-minus signature for the metric. This property goes under the name of {\it microcausality}.

As discussed at length in \cite{Dubovsky:2007ac}, on the one hand microcausality is tied to Lorentz-invariance, but, on the other hand, since it is an operator statement, it must hold on all possible states of the theory, including Lorentz-breaking ones. Now, it so happens that the only Lorentz invariant state of a relativistic QFT is the vacuum. So, what are the implications of the above operator statement when we evaluate it on a state that is not the vacuum? The obvious answer is that certain correlation functions, in particular the retarded and advanced ones, must vanish outside the light-cone, regardless of the state of the system. In particular, for a generic state, such correlation functions will {\it not} be Lorentz-invariant, but they will still obey relativistic causality.

Such considerations were recently leveraged in \cite{Creminelli:2022onn, Heller:2022ejw, Creminelli:2024lhd} in order to derive analytic properties of Fourier-space two-point functions in homogeneous and isotropic Lorentz-breaking states, with the goal of deriving sum rules and positivity bounds for the effective theories describing the low-energy physics of those states. Our approach and goal are similar, but we find it useful, as we will discuss, to consider correlation functions in a mixed $(t, \vec k \, )$ representation.\footnote{The analyticity of correlators in the $(t, \vec k \, )$ representation was independently noticed in \cite{Gavassino:2023mad}.} This, among other things, has the advantage of allowing us to apply our results to cosmological correlators. Another advantage is that the analyticity and boundedness conditions in the mixed $(t, \vec k \, )$ representation are necessary and sufficient for microcausality, whereas analyticity of retarded correlators in $(\omega, \vec k \, )$ is only a necessary condition, unless supplemented by a different boundedness condition, as we will show. 

In particular, we will consider generic spatially translationally and rotationally invariant states in a generic QFT that obeys microcausality. The states can be pure, such as the ground state of a superfluid, or mixed, such as the equilibrium, finite-temperature state of a normal fluid. We will never use Lorentz invariance---since our states break it---and so in a sense it is not even needed that the underlying QFT be Lorentz invariant, provided it obeys microcausality. 
By going to {\it spatial} Fourier space, $\vec x \to \vec k$, but retaining time instead of frequency as a variable, we will show that certain correlation functions $\tilde{G}(t, \vec k \, )$ must be

\begin{enumerate}
\item  analytic in $\vec k$ in all of $\mathbb{C}^3$, or, using isotropy, analytic in the whole $|\vec k|^2$ complex plane;
\item  bounded by
\be
\big | \tilde{G}(t, \vec k \, ) \big| < f\big( \vec k \, \big) \, e^{\vert{\rm Im} ( {\vec k } \, ) \vert \,  t }  \qquad   \forall \vec k \in \mathbb{C}^3 \; ,
\ee
where $f$ is a certain non-exponential factor.
\end{enumerate}

We will first discuss in detail how these properties arise, their relation to analyticity properties in $(\omega,\vec k)$ space and how they are satisfied in Lorentz-invariant theories (Secs.~\ref{general}--\ref{LorentzInv}).
We will then see how these properties are obeyed in a number of physical, Lorentz-breaking situations (Secs.~\ref{superfluids}--\ref{cosmology}). Sometimes, the analyticity property is obeyed thanks to quite nontrivial cancellations, including, in particular, for cosmological correlators in de Sitter space and in slow-roll inflationary theories (Sec.~\ref{cosmology}). In the latter case, exponential boundedness will be obeyed within the EFT regime only if a positivity condition on EFT coefficients is satisfied.

We will also discuss how our results do imply that microcausality can be tested within a Lorentz-breaking low-energy effective field theory, contrary to some claims in the literature (Sec.~\ref{microEFT}). In particular, we find that for an excitation that becomes stable at low momenta, the exponential boundedness criterion implies that, not surprisingly, the group velocity at the frequencies of interest must be subluminal (Sec.~\ref{group}). If, on the other hand, the excitation in question has a sizable decay rate, the group velocity can exceed the speed of light, but only up to a certain degree. Nowhere in the derivation of these results does the so-called front velocity play an important role. Besides clarifying the emergence of the subluminality criterion and its detailed form from first principles, we will also show how microcausality can give strictly stronger bounds that cannot be derived by subluminality of group velocity (Sec.~\ref{loop}).

\vspace{.5cm}
\noindent
{\it Notation and conventions.} We use $\hbar = c = 1$ units and the mostly minus signature for the spacetime metric (except Sec.~\ref{cosmology}). Since in most of the paper we do not use Lorentz invariance or covariance, by $k^2$ we will denote the square of the momentum $3$-vector, $k^2 = \vec k \cdot \vec k$, rather than its relativistic counterpart $k_\mu k^\mu$. We will be working with complex vectors $\vec k \in {\mathbb C}^3$ and thus complex $k^2 \in {\mathbb C}$. It will be useful to have a different notation for the absolute value of a complex number $z$ and the norm of a (real or complex) vector $\vec v$. For the former we write $|z|$, and for the latter $\norm{\vec v}$. 
We use $\hat G$ to denote correlators in $(\omega,\vec{k})$ space, while $\tilde G$ will denote correlators in $(t,\vec{k})$ space.

\section{The general statement}\label{general}
For a QFT obeying microcausality, consider a spatially homogeneous and isotropic state, pure or mixed, and a real $3$-scalar local operator $\phi(t, \vec x)$. This could be a `fundamental' field, that is one of the fields appearing in the path integral formulation of the theory, or a more general composite operator. Microcausality implies that the Green's function
\be
G_c(t, \vec x) \equiv \langle \, [\phi(t, \vec x), \phi(0)] \, \rangle
\label{Gcdef}
\ee 
vanishes outside the light-cone. Here, `$c$' stands for `commutator', and the average $\langle \dots \rangle$ stands for the expectation value $\langle\Psi| \dots |\Psi \rangle$ in the case of a pure state $|\Psi \rangle$, or for the trace ${\rm tr} \big( \rho \dots \big)$ in the case of a mixed state with density matrix $\rho$. We are considering this particular two-point function, but everything we say below applies equally well to the retarded and advanced two-point functions
\be 
G_R(t, \vec x)= \theta(t) G_c(t, \vec x) \; , \qquad G_A(t, \vec x) = \theta(-t) G_c(-t, \vec x)
\ee
as well. In fact, these also vanish for $t<0$ and $t>0$, respectively, but these additional properties will not play an essential 
role in our analysis. We will collect some results on commutator Green's function in QFT in Appendix~\ref{app:generalities}.

Now, the vanishing of $G_c$ outside the lightcone means that, as a function of $\vec x$ at fixed $t$, $G_c$ has compact support, in a ball of radius $R= |t|$. It is a general result of distribution theory and complex analysis (Paley-Wiener theorem, see e.g.~\cite{Stein}) that a function is of compact support \emph{if and only if} its Fourier transform is analytic and exponentially bounded. 

More precisely, if an $L^1$ function $f(\vec x \, )$ has compact support of radius $R$, its Fourier transform $\tilde f(\vec k \,)$ is, for all complex $\vec k$'s, analytic in the vector $\vec k$  and bounded by
\be
|\tilde f(\vec k)|< C e^{\norm{\Im \, \vec k}R}
\ee
for some constant $C$.
If instead of a function we have a tempered distribution $\varphi(\vec x \,)$ of compact support\footnote{A tempered distribution of compact support ${\cal A}$ is a tempered distribution that yields zero if applied to all test functions whose support does not intersect ${\cal A}$.}, 
then its Fourier transform $\tilde \varphi(\vec k\, )$ is still analytic for all complex $\vec k$'s, but the boundedness condition involves a more general non-exponential factor:
\be
|\tilde \varphi(\vec k)| < C (D + \norm{\vec k})^N e^{\norm{\Im \, \vec k} R} \; ,
\ee
for some positive constants $C$, $D$, and $N$, and for all complex $\vec k$'s.

Let us apply these results to our case. Our $ G_c(t, \vec x) $ is a distribution\footnote{We consider here distributions in $\vec{x}$ for fixed $t$. Regarding the $t$ dependence, we will only discuss the case of ordinary functions of $t$, since all the examples we present fall within this class for $t\neq 0$. For the more general case see Appendix~\ref{frequency space}.} with support in a ball of radius $R = |t|$, and so its {\it spatial } Fourier transform $\tilde G_c(t, \vec k \, )$ is, for all complex~$\vec k$'s,
\begin{enumerate}
\item analytic in the complex vector $\vec k$;
\item bounded  by  
\be \label{exp bound vec k}
|\tilde G_c(t, \vec k \, )| < C (D + \norm{\vec k})^N e^{\norm{ {\Im} \, \vec k} \cdot |t|}
\ee
for some positive constants $C$, $D$, and $N$.
\end{enumerate}
The Paley-Wiener theorem ensure that the converse is also true: if $\tilde G_c$ has these properties, then it is the Fourier transform of a distribution with compact support of radius $|t|$.

These are the properties that we will check over and over in the rest of the paper. Before we do so, a few comments are in order
\begin{itemize}
\item
Since we are dealing with a scalar operator and with a rotationally invariant state, our $G_c$ can only depend on $\vec x$ through the combination $|\vec x|^2$. Similarly, its Fourier transform can only depend on $\vec k$ through the combination
\be
k^2 \equiv \vec k \cdot \vec k \; ,
\ee
which is analytic in the complex vector $\vec k$.
So, the analyticity condition above simply boils down to the requirement that $\tilde G_c$ be analytic in the variable $k^2$, for all complex $k^2$'s. Similarly, for an analytic function of $k^2$, the exponential boundedness condition can be rewritten
as
\be
|\tilde G_c(t, \vec k \, )| < C (D + | k | )^N e^{|{\Im}\,  k | \cdot |t|} \; , \qquad \forall  k \in \mathbb{C}
\ee 
where $k$ is simply either of the two complex square roots of $k^2$:
\be
k \equiv \sqrt{k^2} .
\ee
We prove this in Appendix \ref{k vs k}.
From now on, we will use these simpler criteria when checking analyticity and exponential boundedness.

\item
On a related note, we are considering a single operator $\phi$ that is a scalar under rotations (but not necessarily under Lorentz transformations), and a rotationally invariant state. This is just for simplicity: our analysis can be extended to mixed correlation functions of more general operators, in general representations of rotations, and to states that are invariant under spatial translations but break rotations (in which case the comment above about $k^2$ does not apply). 
In fact, in Secs.~\ref{solids} and \ref{hydro} we will consider operators that transform as vectors of the rotation group.

\item
In the Paley-Wiener theorem, the non-exponential prefactors in the boundedness property involve `constants' that are constant only as far as their $\vec k$-dependence is concerned. In our case we have another variable, $t$, which measures the radius of the support of $G_c$. Since in the following we will be taking the large $t$ limit, it is important to understand whether those non-exponential-in-$\vec k$ prefactors can in fact have an exponential dependence on $t$. It seems to us that this can only happen if the state under consideration is unstable. In what follows we will assume these coefficients to grow slower than an exponential.

\item
We have never assumed invariance under {\it time}-translations. Thanks to this, we can apply the conditions above to cosmological correlation functions, assuming microcausality holds there too. In fact, it is an interesting open question to what extent microcausality should apply to cases with dynamical gravity. More on this below.

\item
If one does assume invariance under time-translations, one can go to Fourier space for time as well and rephrase our conditions of analyticity in $\vec k$ space in terms of $\hat G_c(\omega, \vec k)$. However, we found this representation more difficult to use, essentially because, as a function of $\omega$, $\hat G_c$ is in general a distribution rather than a function. This makes checking analyticity in $\vec k$ substantially more difficult. We give some more details on this in Appendix \ref{frequency space}. 
This form of analyticity should be distinguished from the analyticity properties in $(\omega, \vec k)$ space of retarded correlation functions, which we will now discuss.

\end{itemize}

\section{Microcausality in frequency vs time domain}

This work is not the first to ask how microcausality is encoded in analytic properties of correlation functions. Previous works focused mainly on the frequency space representation of retarded correlators~\cite{Creminelli:2022onn,Heller:2022ejw,Creminelli:2024lhd} (for a detailed discussion of causality and analyticity in coordinate space see e.g.~\cite{Bogolubov:1990ask,Kravchuk:2021kwe}). 
In frequency space, microcausality implies that retarded correlation functions are analytic functions of complexified frequency and momentum in the region ${\rm Im}(\omega)> \vert {\rm Im}(\vec{k})\vert$.\footnote{As we will review in the following, this property is derived assuming a stability condition, namely that correlators do not grow exponentially with time (see also the comment in the previous section).} A complementary condition can be derived for advanced correlators, in the region ${\rm Im}(\omega)< - \vert {\rm Im}(\vec{k})\vert$. Notice that the analyticity properties of the commutator in frequency space are more subtle, since this can be written as a linear combination of advanced and retarded correlators, and the respective regions of analyticity have zero overlap (see however the discussion in Appendix~\ref{frequency space}). Many interesting consequences can be derived from analyticity properties of retarded Green's function, even in Lorentz-breaking states, as discussed at length in~\cite{Creminelli:2022onn,Heller:2022ejw,Creminelli:2024lhd}.

These properties are necessary conditions for microcausality, but in fact they are not sufficient. 
The easiest way to see this is by looking at some counterexamples:
\begin{itemize}
\item[A)] Consider a retarded correlator that is constant in a time-independent support in~$\vec x$: $G_R(t,\vec{x})=\theta(t)  \theta(R- \vert\vec{x}\vert) $.

\item[B)] Consider the retarded correlator: $G_R(t,\vec{x})= \, \theta(t) \, e^{-\vert\vec{x}\vert^2/R^2}\; .$
\end{itemize}
In both cases it is easy to check that the Fourier transform in space and time is analytic in the region ${\rm Im}(\omega)>0$, and therefore also in the region of analyticity of causal retarded correlators ${\rm Im}(\omega)> \vert {\rm Im}(\vec{k})\vert$. On the other hand, neither correlator  vanishes outside the lightcone defined by $\vert \vec{x} \vert = t$. Therefore, this form of analyticity is not sufficient to ensure microcausality. 

In fact, under the same stability condition mentioned previously and the assumption that the equal time commutator vanishes,\footnote{The assumption that the equal time commutator vanishes, $G_c(t=0,\vec{x})=0$, is always valid in Lorentz invariant theories when the Kh\"allen-Lehmann spectral representation is convergent, as follows from eq.\eqref{eq:spectralrep} by taking the $t\to 0$ limit. In general, Lorentzian correlators require smearing in time, see e.g.~\cite{Witten:2023qsv} (we thank P. Creminelli and B. Salehian for discussions on this point). Our approach based on the mixed $(t,\vec{k})$ representation will be valid in either case.} microcausality implies some additional properties of retarded correlators:
\be \label{eq:vanish}
\begin{split}
&\hat{G}_R(\omega,\vec{k})\to 0 \qquad {\rm for} \qquad {\rm Im}\,\omega \to + \infty \; , \quad \vec k \in {\mathbb C}^3 \;\;  {\rm fixed} \; ,\\
& \hat{G}_R(\omega,\vec{k}) \qquad {\rm grows  \;at \; most\; polynomially \; for}\qquad ({\rm Im}\,\omega - \norm{ {\rm Im}\,\vec{k}} )\to + \infty \; .
\end{split}
\ee
Similar properties hold for advanced correlators.
To see how they arise, let us write the frequency space retarded correlator in terms of the commutator Green's function in the mixed $(t, \vec k \,)$ representation. We have 
\be
\left\vert \hat{G}_R(\omega,\vec{k}) \right\vert \leq \int_0^\infty dt \vert \C(t,\vec{k}) \vert \, \vert e^{i\omega t}\vert \leq  \int_0^\infty dt \,C(t,\vec{k})\, e^{-({\rm Im}\,\omega - \norm {{\rm Im}\,\vec{k} } ) t} \; ,
\ee
where the first inequality follows from the definition, while in the second inequality we used the Paley-Wiener condition derived from microcausality and discussed in section~\ref{general}. As long as $C(t,\vec{k})$ grows slower than an exponential in time and is regular at $t=0$, the last term can be identified as the Laplace transform of $C(t,\vec{k})$ with $s\equiv({\rm Im}\,\omega - \norm{ {\rm Im}\,\vec{k}} )$. It is a standard result of analysis that this converges for $s>0$, and goes to zero in the limit $s\to +\infty$ if we take the limit with fixed $\vec k \in {\mathbb C}^3$.\footnote{If $C(t,\vec{k})$ is polynomially bounded in time, one can similarly obtain a stronger power-law bound on the behaviour of the retarded correlator for ${\rm Im}\,\omega \to + \infty$.} More generally, it can grow at most polynomially for $s\to \infty$, since $C(t,\vec{k})$ is a polynomial in $\vec k$ (see eq.~\eqref{exp bound vec k}). 

The asymptotic properties~\eqref{eq:vanish} are sufficient to exclude the counterexamples discussed above. An argument of Salehian and Creminelli shows that, together with analyticity, they are in fact sufficient conditions for microcausality.\footnote{We thank B. Salehian and P. Creminelli for sharing with us a preliminary version of their result, and for useful discussions on these conditions.}

\section{The Lorentz invariant case }\label{LorentzInv}

We can now consider some specific examples. As a warmup, we start with the Lorentz invariant case.
Consider the commutator Green's function for a free scalar field $\phi$ of mass $M\geq0$ in its Poincar\'e invariant vacuum. A straightforward computation, as detailed in Appendix~\ref{app:generalities}, yields
\be
\tilde G_c (t, \vec k \, ) = -i\,\frac{\sin\big(\omega_k \, t\big)}{\omega_k}  \; , \qquad \omega_k \equiv \sqrt{k^2+M^2} \; .
\ee
Because of the square-root structure, $\omega_k$ is not analytic in $k^2$. However, by Taylor-expanding  $\sin(\omega_k \,t)$, which is an {\it odd} analytic function of $\omega_k$, we get that $\tilde G_c$ is an {\it even} analytic function of $\omega_k$. $\tilde G_c$ is thus analytic in $\omega^2_k = k^2 +M^2$, and therefore analytic in~$k^2$, for all $k^2 \in  \mathbb{C}$, as needed. 
 
The exponential boundedness is more delicate to check. Ignoring non-exponential prefactors, we have
 \be
 \tilde G_c (t, \vec k \, ) \sim e^{\pm i \omega_k t} ,
 \ee
which for real $\vec k$ oscillates at frequency $\omega_k$, that is, {\it faster} than $k$. This is the usual statement that the phase velocity $v_{\rm ph} = \omega_k/k$ of a massive particle is always {\it superluminal}.\footnote{At this point in our analysis there is no obvious connection between microcausality and the group velocity $v_g= d \omega_k/dk$, and so we will postpone discussing such a connection until later.}
However, we have
\be
| \tilde G_c (t, \vec k \, ) |  \lesssim e^{ | {\Im}\, \omega_k| \cdot |t|} \; ,
\ee
and it so happens that, even though $\omega_k$ is always bigger than $k$ for real $\vec k$, their imaginary parts obey the opposite inequality,
\be
|{\Im} (\,\omega_k) | \le |{\Im} \, k|  \; ,
\ee 
for all complex $\vec k$'s
(we show this in Appendix \ref{imaginary}). And so, our Green's function is analytic in $k^2$ and, up to non-exponential factors, bounded by
\be \label{exp bound relativistic}
| \tilde G_c (t, \vec k \, ) |  \lesssim e^{ | {\Im} \,k| \cdot |t| } \; ,
\ee 
as expected.

We can now combine what we just learned for a free massive scalar field, with the spectral decomposition of the vacuum two-point function of a {\it generic} scalar operator $\phi(x)$ in a {\it generic} interacting relativistic QFT. At the level of $T$-ordered correlation functions, the K\"allen-Lehman decomposition reads
\be
\langle T \phi(t, \vec x) \phi(0) \rangle = \int_0^{\infty} d\mu^2 \rho(\mu^2) \, G_F(t, \vec x; \mu^2) \; ,
\ee
where the spectral density $\rho(\mu^2)$ depends on the spectrum of the theory and on the operator $\phi$ under consideration, and $G_F(t, \vec x; \mu^2)$ is simply the Feynman propagator for a free scalar field of mass $\mu$. 
If instead of the $T$-ordered correlation function, we are interested in the expectation value of the commutator, we get exactly the same spectral decomposition, with the same $\rho(\mu^2)$, but with $G_F$ replaced by the free massive scalar $G_c$ that we analyzed above, with mass $\mu$. That is, in $(t, \vec k \, )$ space we have
\be\label{eq:spectralrep}
\tilde G_c (t, \vec k \, ) =  -i \int_0^{\infty} d\mu^2 \rho(\mu^2) \, \frac{\sin\big(\omega_k(\mu^2) \, t\big)}{\omega_k(\mu^2)} \; , \qquad \omega_k(\mu^2) \equiv \sqrt{k^2+\mu^2} \; .
\ee
As we proved above, for all $\mu^2$'s the integrand is analytic in $k^2$ and exponentially bounded as in \eqref{exp bound relativistic}. Assuming the integral in $\mu^2$ converges, the resulting $\tilde G_c (t, \vec k \, )$, as a function of $\vec k$, obeys the same properties.

This shows that, not surprisingly, the conditions that follow from microcausality are automatically obeyed by generic scalar two-point functions in the Poincar\'e-invariant vacuum of a general relativistic QFT. We implicitly made use of Lorentz invariance, because we used, crucially, that the spectral density is only a function of the invariant mass $\mu^2$, rather than separately of $\omega$ and $\vec k$. As we will see below, without this property checking the analyticity of $G_c$ in specific examples can be quite involved, especially beyond tree-level.

\section{Microcausality in effective field theory} \label{microEFT}
An important question is whether microcausality can be checked within a Lorentz-breaking low-energy effective field theory
without knowing what the UV completion is. The general counterargument is that resolving the lightcone involves having access to arbitrarily short distances and times, that is, arbitrarily high momenta and energies, which is clearly something impossible in EFT (see for example \cite{Hollowood:2007kt,Hollowood:2007ku,Hollowood:2015elj}).

However, our analyticity and exponential boundedness constraints apply for all values of $\vec k$ and $t$, and EFT allows us to systematically compute the low-$\vec k$, large $t$ behavior of correlators, to any desired order. In particular, analyticity at $\vec k = 0$ is something that can be checked within EFT: as is well known \cite{Manohar:2018aog}, for scattering amplitudes and $T$-ordered correlation functions, the non-analyticities in external momenta for $\omega$ and $\vec k$ within the regime of validity of the EFT are {\it completely} captured by the low-energy effective field theory; when one goes from $T$-ordered correlations functions to the expectation values of commutators, and from frequency space to time, our claim is that, if microcausality holds, such non-analyticities should disappear.

As to exponential boundedness, we would like to have a simple check that does not require us to know explicitly the prefactor in \eqref{exp bound vec k}. We can then consider limits in which the exponent in
\be
e^{\norm{ {\Im} \, \vec k} \cdot |t|}
\ee
is much larger than one.  Within an EFT, we can achieve such a limit by taking
\be
\label{ktlarge}
k \ll \Lambda_k \; , \quad t \gg \Lambda_\omega^{-1} \; , \quad \mbox{with }k \, t \gg 1 \; ,
\ee 
where $\Lambda_k$ and $\Lambda_\omega$ are the momentum and energy UV cutoffs of the effective theory.

We thus reach the conclusion that we can reliably check whether a low-energy EFT is compatible with microcausality, while still being agnostic about the high momentum, high energy behavior of correlators.
%
\subsection{Simple examples}\label{simpleExamples}

To illustrate the previous discussion on microcausality in effective field theory, we can look at the case of a non-relativistic free field theory.
Consider a relativistic free scalar $\phi(t,\vec{x})$ with mass $m$. In a non-relativistic approximation, $k \ll m$, this can be described in terms of the mode decomposition
\begin{align}
\hat{\phi}(t,\vec{x}) & = \hat{\psi}(t,\vec{x}) e^{-im t }+ \hat{\psi}^\dagger(t,\vec{x}) e^{im t } \\
\hat{\psi}(t,\vec{x}) & \simeq \int \dfrac{d^3 k}{(2\pi)^3 \sqrt{2 m}} \, \hat{a}_{\vec k} \, e^{- i E_k t} e^{i \vec{k}\cdot \vec{x}} \; , \qquad  E_k \simeq \frac{k^2}{2m} \; ,
\end{align}
with a low-energy  effective action for $\psi$
\begin{equation}
S_{\rm NREFT}= m\int \psi^*(t,\vec{x}) \left( i \partial_t + \dfrac{\nabla^2}{2m} \right)\psi(t,\vec{x})  \, d^4x.
\end{equation}
The commutator $G_c=\langle [\psi, \psi^\dagger] \rangle$ is simply 
\begin{equation}
 \C(t, \vec k) = \dfrac{1}{2m} e^{-i E_k \, t }.
\end{equation}

Analyticity in $k^2$ is trivially satisfied, because now $E_k$ is analytic in $k^2$. On the other hand, exponential boundedness (and thus microcausality) is satisfied only as long as $|\Im{\, k^2}| /2m \leq |\Im \, k|$, which is certainly the case in  the non-relativistic approximation, $k \ll m$. On the other hand,  had we not known about the relativistic UV completion of the non-relativistic effective theory,  we could have used exponential boundedness to argue---correctly---that the momentum cut-off of the EFT should be at most $\Lambda_{\rm UV}\sim m$.

As another simple example, we can consider the derivatively coupled relativistic scalar of Ref.~\cite{Adams:2006sv},
\be
{\cal L} = \sfrac12 (\partial \phi)^2 + \sfrac14 \alpha (\partial \phi)^4 + \dots ,
\ee
where the dots in the effective Lagrangian stand for higher derivative interactions, involving higher powers of $\partial_\mu \phi$ and/or higher derivatives of $\phi$. It well known that $S$-matrix unitarity and analyticity requires $\alpha$ to be positive.

If we expand the action around a linear-in-time solution, $\phi(x) = C \, t +\pi(x)$, with a perturbatively small constant $C$,
for the fluctuation $\pi(x)$ we get a Lorentz-breaking quadratic Lagrangian,
\begin{align}
{\cal L} &\to \sfrac12 (1+ \alpha C^2) (\partial \pi)^2 +  \sfrac12 \alpha C^2 \dot \pi^2 + \dots \\
& = \sfrac12 (1+ 3 \alpha C^2) \dot \pi^2 -  \sfrac12  (1+ \alpha C^2)  (\nabla \pi)^2 + \dots \\
& \to \sfrac12  \dot \pi^2 -  \sfrac12  (1- 2\alpha C^2)  (\nabla \pi)^2 + \dots \; ,
\end{align}
where in the last step we changed the normalization of $\pi$, and we expanded to first order in $\alpha C^2$.

At this order, the commutator 
$G_c=\langle [\pi, \pi] \rangle$  is simply
\be
 \C(t, \vec k) = -i \frac{\sin\left(\omega_k \, t \right)}{\omega_k}  \; , \qquad \omega_k \simeq k (1- \sfrac12 \alpha C^2) \; .
 \ee
 Analyticity in $k^2$ now works exactly like in the relativistic case. However, exponential boundedness, clearly requires that $\alpha$ be positive, in agreement with the standard results.
 
The same of course holds for any free field whose Lagrangian starts, in a derivative expansion, as
\be
{\cal L}= \sfrac12  \dot \pi^2 -  \sfrac12  c_s^2  (\nabla \pi)^2 + \dots \; ,
\ee
 where $c_s$ is an effective ``sound speed". Analyticity is automatic, while exponential boundedness, not surprisingly,
 requires
 \be
 c_s \le 1 \; .
 \ee

\subsection{The emergence of the group velocity}\label{group}
Sometimes, the causal response of a system is dominated, at least in a certain range of times and wavenumbers and within some approximations, by a single excitation with definite dispersion law $\omega = \omega_k$. For example, if for some real $\vec k$ the Fourier-space retarded Green's function $\tilde G_R(\omega, \vec k \,)$ has a pole at a relatively low $\omega = \omega_k$ and all other non-analyticities  are at much higher $\omega$'s, then such a pole will dominate the response of the system at that wavenumber for times of order $\omega_k^{-1}$ and longer. 

When such an approximation holds, ignoring non-exponential factors, one has
\be
\tilde G_R(t, \vec k \, ) \sim e^{- i \omega_k \, t} \; .
\ee 
What are the implications of microcausality in this case?
Without  knowledge of the $\vec k$-dependence of the residue and thus of the non-exponential prefactor, we cannot check analyticity in $\vec k$, but we can at least check exponential boundedness. In order to obey it, we should have
\be
\Im \, \omega_k  \le |\Im \, k| \; .
\ee

Consider then such a constraint for $k$ close to the real axis: {\it If} the excitation's frequency is real for real $\vec k$, that is, if our excitation is exactly stable, we have
\be \label{group velocity imaginary part}
\Im \, \omega_k \big|_{k \to \mathbb{R}^+} \simeq \frac{d \omega_k}{dk} \, \Im \, k \; ,
\ee
and so our constraint is obeyed for those $\vec k$'s if and only if the group velocity is subluminal,
\be
 |v_g(k)|  \le 1 \; ,\qquad v_g(k) \equiv  \frac{d \omega_k}{dk} \; .
\ee

On the other hand, if for real $\vec k$ our excitation has a width $\Gamma_k > 0$,
\be
\Im \, \omega_k = - \frac{\Gamma_k}{2} 	\; , \qquad k = \mathbb{R}^+ \; ,
\ee
then \eqref{group velocity imaginary part} gets modified as
\be \label{omega Gamma}
\Im \, \omega_k \big|_{k \to \mathbb{R}^+} =  v_g \, \Im \, k - \frac{\Gamma_k}{2} \; ,
\ee
where now the group velocity is  defined as
\be
v_g (k)= \frac{d \, \Re \, \omega_k}{dk}  \; .
\ee
The exponential boundedness constraint now becomes
\be
v_g(k) \, \Im \, k - \frac{\Gamma_k}{2} \le |\Im \, k| \; .
\ee
Although $v_g$ and $\Im \, k$ can in principle have either sign, such a constraint is stronger for positive $v_g \cdot \Im \, k$. Without loss of generality, we can then rewrite it as
\be \label{vg Gamma}
|v_g(k)| \le 1 +  \frac{\Gamma_k}{2 |\Im \, k|}  \; .
\ee
That is, in this case the group velocity does not have to be subluminal. 

That group velocities can be superluminal is old news---see for example \cite{Jackson:1998nia} and references therein.
What our analysis shows is the direct relationship of this statement with the width of the excitation in question.
There appears to be no bound, since we derived the inequality~\eqref{vg Gamma} in the limit of infinitesimal $\Im \, k$, so that the right side is effectively infinite. 

However, in the case of a narrow resonance, $\Gamma_k \ll \Re \, \omega_k$, this conclusion is premature. The reason is that, if $\omega_k$ varies smoothly as a function of $k$ in the region of interest, the relationship \eqref{omega Gamma} is approximately correct for non-infinitesimal values of $\Im \, k$, as long as these are not too large. How large is too large? It depends on the dispersion relation at hand, but the important point is that the constraint \eqref{vg Gamma} can give a non-trivial constraint on the group velocity. 

As an example, consider phonons in a superfluid. At low momenta, they have a dispersion relation that can be parametrized as \cite{Landau9}\footnote{We are considering superfluids such that the term of order $k^2$ in brackets is positive. When it is negative, low-energy phonons are exactly stable \cite{Landau9}, and our previous analysis applies.}
\be
\Re \, \omega_k = c_s k \, \big[1 + \sfrac{1}{3}  (k / \kappa)^2 + {O}(k/\kappa)^4 \big] \; , \qquad  \Gamma \simeq  c_s k \cdot \alpha  (k/\kappa)^4  \; ,
\ee
where $c_s$ is the sound speed in the $k \to 0$ limit, $\kappa$ is a microscopic momentum scale that determines the validity of the derivative expansion, $k \ll \kappa$,  and $\alpha$ is a (small) numerical parameter. The width comes from the decay of a phonon into two lower energy ones, which is more and more suppressed the lower $k$ is, since phonons are derivatively coupled. The group velocity is 
\be
v_g(k) =  c_s [1 + (k / \kappa)^2 + \dots] \; ,
\ee
and the approximation \eqref{omega Gamma} is valid as long as 
\be
\Im \, k \ll \kappa \; ,
\ee
regardless of the value of $\Re \, k$, provided this is also much smaller than $\kappa$. Then, taking $\Im \, k$ of order $\Re \, k$ or larger in eq.~\eqref{vg Gamma} we simply get
\be
v_g(k) \simeq c_s [1 + (k / \kappa)^2 \big ] \le 1 + O(k/\kappa)^4 \; .
\ee
That is: within the regime of validity of the low-energy phonon effective theory, the group velocity computed to order $k^2$ cannot be superluminal; in particular, $c_s$ must be {\it strictly} subluminal:
\be
v_g(k \ll \kappa) \le 1 \; , \qquad c_s<1 \; .
\ee
Such a conclusion  will perhaps not surprise anybody, but here we derived it with a modicum of rigor.
As we will see in Sec.~\ref{loop}, for this specific example the analysis is cleaner and more powerful at the level of correlation functions.

\section{Superfluids}\label{superfluids}

We now consider zero-temperature superfluids, which are the simplest systems that break Lorentz boosts spontaneously.  In a relativistic QFT with a conserved $U(1)$ charge, a superfluid state is a state that {\it i)} has a nonzero density for that charge, and {\em ii)} spontaneously breaks the corresponding $U(1)$ symmetry. All such states break Lorentz boosts, because any nonzero charge density, $\langle J^0 \rangle$, does so.

\subsection{A simple UV complete theory}
A simple relativistic 
UV completion to keep in mind is a finite-density state in the theory of a complex scalar $\Phi$ with $U(1)$ invariant quartic interactions \cite{Hui:2023pxc,Creminelli:2023kze},
\begin{equation}
    S = \int d^4x \, |\partial \Phi|^2 + M^2|\Phi|^2 - \lambda |\Phi|^4 \; ,
\end{equation}
where for simplicity we consider the positive $M^2$ case, so that the $U(1)$ symmetry is spontaneously broken already in the Poincar\'e invariant vacuum. Without loss of generality, we take the expectation value of $\Phi$ aligned with the positive real axis, ${\langle \Phi \rangle \in \mathbb R^+}$.

To study states with finite charge density, we introduce a chemical potential $\mu$. There is a gapless excitation---the superfluid phonon---and a gapped one, with dispersion relations \cite{Hui:2023pxc}
\begin{equation} \label{omega pm}
    \omega_\pm(k) = \sqrt{k^2 + \frac12 \Delta^2\Big( 1 \pm  \sqrt{1 + 4(1-c_s^2){k^2/\Delta^2}} \Big)}     \; ,
\end{equation}
where `$-$' corresponds to the gapless one and `$+$' to the gapped one, and $c_s$ and $\Delta$ are, respectively, the sound speed and gap  for $k \to 0$. Notice the double square root structure, which will make the analyticity-in-$k^2$ check quite nontrivial.

Moreover, the overlaps of the local fields $\Phi$ and $\Phi^*$ with the single particle states also involve double square roots. For simplicity, consider the real local field
\be
\phi_2 = \sqrt{2} \, \Im \, \Phi \; ,
\ee
which interpolates the Goldstone already at vanishing chemical potential. For a single particle state $|\vec k, \pm \rangle$, at tree level one finds \cite{Hui:2023pxc}
\be
Z_{\pm} (k) \equiv \langle \mu | \phi_2(0) | \vec k, \pm \rangle =  \mp \frac{i}{\sqrt{2}} \sqrt{ 1 \pm \dfrac{(1-2 c_s^2)}{\sqrt{1+4 (1-c_s^2) k^2/\Delta^2}}} \; ,
\ee
where $|\mu \rangle$ is the  finite-$\mu$ ground state

Now consider the commutator of $\phi_2$ with itself in the finite-$\mu$ ground state,
\be
G_c(t, \vec x) = \langle \mu | [\phi_2(x), \phi_2(0)] | \mu \rangle  \; .
\ee
Using the K\"all\'en-Lehmann decomposition, we find 
\begin{equation}
    \tilde G_c(t, \vec k \, ) = \sum_{\alpha = \pm}  |Z_\alpha(k)|^2  \times (-i) \, \frac{\sin \left( \omega_\alpha(k) \, t\right)}{\omega_\alpha(k)}   \; .
\end{equation}
To obey microcausality, this should be analytic in $k^2$ and exponentially bounded.

As for analyticity:  the squared absolute value and the $\frac{\sin z}{z}$ structure get rid of the outer square roots in $\omega_\alpha$ and $Z_\alpha$, but we are still left with the inner ones, which, if we consider the two excitations separately, do not cancel. However, when we consider the sum over $\alpha$, the associated non-analyticities do cancel, and we are left with a function that is analytic in $k^2$. A simple way to see how this can happen is to parametrize $\omega_\alpha^2$ and $|Z_\alpha|^2$ as
\be
\omega_\pm ^2 = a \pm b \; , \quad  |Z_\pm|^2 = c \pm 1/{b} \; , \qquad b \equiv \sqrt{1 + 4(1-c_s^2){k^2/\Delta^2}} \; ,
\ee
where $a$ and $c$ are analytic in $k^2$, and for simplicity we are neglecting a common, $k$-independent normalization factor in $Z_\alpha$. Then, Taylor expanding the $\sin$ in $\tilde G_c$ we get a series of terms of the form
\be
\left( c + {1}/{b} \right) \left( a + {b} \right)^{n} + \left( c - {1}/{b} \right)\left( a - {b} \right)^{n} \; , \qquad  n=0,1,2, \dots
\ee
For given $n$, such a combination is even in $b$ and, despite the apparent $1/b$ singularity, which cancels between the `$+$' and `$-$' contributions, analytic in $b$. We thus have analyticity in $b^2$, which implies analyticity in $k^2$!

Notice that the magic only works if we sum over both types of excitations. So, in a sense, the two types of excitations need each other, with precisely tuned properties ($\omega_\alpha$ and $Z_\alpha$) so that microcausality can be respected.
How can this be consistent with effective field theory? After all, at low enough energies there should be a consistent description of the system in which we only keep the gapless excitation and, as we argued in the previous section, we should be able to check microcausality within that effective theory. 
Indeed, the non-analyticity associated with $b$ only shows up at large $k^2$: the branch point of that square root is at $k^2 \gtrsim \Delta^2$  \cite{Hui:2023pxc}. If we restrict our analysis to much smaller $k$'s, as one should in effective field theory, the contribution of each excitation to the commutator is separately analytic in $k^2$.

Exponential boundedness is more tedious to check. It holds separately for the two excitations' contributions, but for the sake of brevity here we omit the proof.

\subsection{The effective theory beyond leading order}\label{loop}

Using the phonon low-energy effective theory, we can compute the one-loop corrections to our correlators, and check whether 
these obey analyticity and exponential boundedness as well---of course for momenta and times within the regime of validity of said effective theory.

We can do this for a generic phonon effective theory, not just for that associated with the UV completion we just described.
The effective action is
\begin{equation} \label{P(X)}
    S = \int d^4x \, P(X) + \mbox{higher derivatives} \; ,
\end{equation}
where $P$ is a smooth function, $X$ is defined as
\begin{equation}
    X = \partial_\mu \psi \partial^\mu \psi ,
\end{equation}
and $\psi(x)$ is a Goldstone field that shifts under the broken $U(1)$ symmetry, $\psi \to \psi +a$.

The ground state at finite chemical potential $\mu$ corresponds to the background solution $\psi= \mu \, t$, and the superfluid phonon ${\pi}(x)$ describes  fluctuations about such a background,
\begin{equation}
    \psi(x) = \mu \, t+ {\pi}(x) \; .
\end{equation}

Then $P(X)$ can be written as
\begin{equation}
    P(X)=P( \mu^2 + 2 \mu \dot{{\pi}} + \dot{{\pi}}^2 - (\nabla {\pi})^2) \; ,
\end{equation}
which we can expand about $X=\mu^2$. Using the canonically normalized field
\begin{equation}
    \pi_c = \sqrt{2P'(\mu^2) + 4 \mu^2 P''(\mu^2)} \, {\pi} \; ,
\end{equation}
where primes denote  derivatives w.r.t.~$X$, we can write the quadratic part of the action as
\begin{equation}
    S_{\rm free} = \frac{1}{2} \int d^4x \,  ( \dot{\pi}_c^2 - c_s^2 \, (\nabla \pi_c)^2  ),
\end{equation}
where
\begin{equation}
    c_s^2 = \frac{P'(\mu^2)}{P'(\mu^2) + 2 \mu^2 P''(\mu^2)} \; .
\end{equation}

Expanding the function $P(X)$ to cubic order in $\pi$, we find the interactions
\begin{equation}
    S_{\rm int} = \int d^4x \, \big[ g_3 \, \dot{\pi}_c \partial_\mu \pi_c \partial^\mu \pi_c + h_3 \, \dot{\pi}_c^3 \big] + {\cal O}(\pi_c^4),
\end{equation}
where
\begin{align}
    g_3 &= \frac{2 P''(\mu^2) \mu }{\left(2P'(\mu^2) + 4 \mu^2 P''(\mu^2) \right)^{\frac{3}{2}}}, \\
     h_3 &= \frac{4 P'''(\mu^2) \mu^3 }{3\left(2P'(\mu^2) + 4 \mu^2 P''(\mu^2) \right)^{\frac{3}{2}}}.
\end{align}
Notice that, in principle, at one loop the two-point functions receives contributions from the quartic interactions as well. However, these do not depend on the external momentum, and are thus completely degenerate with local counterterms, with which eq.~\eqref{P(X)} has to be supplemented anyway. For our purposes, we can thus neglect the quartic interactions.

We spare the reader the computation of  the one loop self-energy diagram, which, because of the lack of manifest Lorentz invariance, is quite tedious.  
The result is
\begin{equation}
    i\Sigma(\omega, \vec k \, ) = i \, \omega^2 \left[A(\omega, k) - B(\omega, k)  \log{\big({c_s^2 {k}^2 - \omega^2} - i\epsilon \big)}  \right],
\end{equation}
where $A$ and $B$ are homogeneous polynomials of fourth-order, and the log has a branch cut on the negative real axis of its argument, that is, for $\omega^2 > c_s^2 k^2$. The polynomial $A$, similarly to the comments above about quartic interactions, is degenerate with higher-derivative local counterterms in the action. Related to this, it depends on the regularization and renormalization prescriptions used in the computation. However, the polynomial $B$ is calculable and, in particular, scheme-independent. It is
\be
B(\omega, k) = \frac{1}{16 \pi^2  c_s^7} \big[b_0 \cdot c_s^4 {k}^4 + b_2 \cdot c_s^2 k^2  \omega^2+ b_4 \cdot \omega^4 \big]\; ,
\ee
with
\begin{align}
     b_0 &= \frac1{40}\big({(49 + 22 c_s^2 + 9 c_s^4) g_3^2 + 2 c_s^2 (11 + 9 c_s^2) g_3 h_3 + 9 c_s^4 h_3^2} \big), \\ 
     b_2 &=- \frac14 {\left((3 + c_s^2)  g_3 + c_s^2 h_3 \right)  \left(g_3 + 3 c_s^2 g_3 + 3 c_s^2 h_3 \right)}, \\
     b_4 &= \frac{1}{8}{(g_3 + 3 c_s^2 g_3 + 3 c_s^2 h_3)^2}.
\end{align}

We checked that the imaginary part of such self-energy reproduces the correct decay rate for on-shell phonons through the optical theorem,
\be
\Gamma = \frac{1}{\omega} \, \textrm{Im} \, \Sigma(\omega,\vec k \,)  \; , \qquad \omega \to c_s k \; ,
\ee
as computed for example in \cite{Hui:2023pxc}.

From the one-loop self-energy, we can compute the one-loop corrected $T$-ordered correlation function of $\pi_c$ in the usual way:
\begin{align}
    G_T(\omega, \vec {k} \,) &= \frac{i}{\omega^2 - c_s^2 k^2 + Q_{c.t.}(\omega, k) + \Sigma(\omega, k) + i \epsilon } \\
    &= \frac{i}{\omega^2 - c_s^2 k^2 +  Q(\omega, k) - \omega^2 B(\omega, k) \log(c_s^2 {k}^2 - \omega^2 - i \epsilon)} \; ,
\end{align}
where $Q_{c.t.}$ is a polynomial of terms of fourth and sixth order, coming from higher derivative corrections  and local counterterms in the effective action, and $Q$ combines $Q_{c.t.}$ with $\omega^2 A(\omega, k)$, and thus has a similar structure to $Q_{c.t.}$, but with different coefficients (which, in the effective theory, are arbitrary anyway).

From $G_T$, we can compute the retarded two-point function simply by changing the $i \epsilon$ prescription: all $\omega$'s should be replaced by $\omega + i \epsilon$. We thus have
\be
G_R(\omega, \vec {k} \,) = \frac{i}{\omega' {}^2 - c_s^2 k^2 +  Q(\omega' , k) - \omega' {}^2 B(\omega', k) \log(c_s^2 {k}^2 - \omega'{}^2)} \; , \qquad \omega' = \omega+ i \epsilon \; .
\ee
Now, a {\it necessary} condition for this to obey microcausality is that it be analytic in $\omega$ and $k^2$ for
\be \label{required}
\Im \, \omega > |\Im \, k| \; .
\ee
Recall that $Q$ and $B$ are polynomials. The potential sources of non-analyticity thus are the cut of the log and the zeroes of the denominator. As to the former, it happens to be located at
\be
(\omega+i \epsilon)^2 = c_s^2 (k^2 + m^2) \; , \qquad m^2 \in \mathbb{R}^+\; ,
\ee
where $m^2$ is a coordinate along the cut. We can rewrite this as the dispersion relation for a relativistic massive particle,
\be
\Omega^2 = k^2 + m^2 \; , \qquad \Omega \equiv \frac{\omega + i \epsilon}{c_s}  \; ,
\ee
which, as proved in Appendix \ref{imaginary}, always has $\Im \, \Omega \le |\Im \, k|$. We thus see that all points on the branch cut
have
\be
\Im \, \omega \le c_s |\Im \, k| - \epsilon < c_s |\Im \, k| \; ,
\ee
which clearly does not intersect the desired analyticity region \eqref{required}, provided the sound speed is subluminal, of course.

As to the poles of $G_R$, we must remember that we are working in effective field theory, which is reliable only at low momenta and frequencies. This means, in particular, that we can only consider the low-energy poles, and not all the spurious poles that we get from the high-order polynomials at the denominator. In other words, if at tree level we only have the two poles
\be
\omega = \pm c_s k - i \epsilon \qquad \mbox{(leading order),}
\ee
at one-loop order we can only trust the perturbative corrections to these two poles, and we must disregard any new high-energy pole that arises because we changed the order of the polynomial at the denominator. And so, the leading correction to the poles comes not from our one-loop computation, but from the uncalculable quartic terms in $Q(\omega, k)$,
\be \label{poles loop}
\omega \simeq \pm c_s k \big(1  + \alpha k^2 \big)  - i \epsilon \; \qquad \alpha \equiv -\frac{Q(c_s k, k)}{2c^2_s k^4} \big|_{k \to 0} \qquad \mbox{(next to leading order).}
\ee
The imaginary parts obey
\be \label{Im omega one loop}
\Im \, \omega \simeq \pm c_s \Im \, k \big(1 +  \alpha(3  \Re^2  k - \Im^2 k) \big) \; .
\ee
Assuming $\alpha$ is positive, if we vary $k \in \mathbb{C}$ while keeping $|k|$ fixed and within the effective theory, say $|k| \lesssim \Lambda_{\rm UV}$,
 the ratio $\Im \, \omega / |\Im \, k| $ is maximized for $k$ close to the real axis:
\be
\max \frac{\Im \, \omega}{ |\Im \, k| }  = c_s \big(1 +  3 \alpha \,  |  k|^2 \big)  \; .
\ee
For this to be smaller than one, we need
\be
\alpha < \frac{1-c_s}{3 c_s |k|^2} \; .
\ee
Moving $|k|$ all the way to the UV cutoff of the effective theory, this can be interpreted as a parametric upper bound on the scale that controls the higher derivative corrections in \eqref{poles loop}: 
\be \label{bound alpha}
\alpha \, \Lambda_{\rm UV}^{2}  \lesssim \frac{1-c_s}{c_s} \; .
\ee

For $c_s$ of order one and not too close to one, the statement is  that, not surprisingly, $\alpha$ cannot be much larger than unity in units of the UV cutoff. However, there are two interesting limiting cases:
\begin{enumerate}
\item
For $c_s$ approaching the speed of light, the bound becomes much stricter: $\alpha$, in such units, must go to zero at least as
\be
\alpha \, \Lambda_{\rm UV}^{2}  \lesssim {1-c_s} \; , \qquad c_s \to 1 \; .
\ee 
\item
In the non-relativistic limit, where $c_s$ is much smaller than the speed of light, the bound becomes looser, but is still nontrivial
\be
\alpha \, \Lambda_{\rm UV}^{2}  \lesssim \frac{1}{c_s} \; , \qquad c_s \ll 1 \; .
\ee
\end{enumerate}

We can actually check these statements in the UV completion discussed above: expanding the phonon's dispersion relation $\omega_-(k)$ \eqref{omega pm} up to $k^3$ order, we get
\be
\alpha \simeq \frac{(1-c^2_s)^2}{2 c_s^2 \Delta^2} \; ,
\ee
where $\Delta$ is the gap, which can be thought as the {\em energy} cutoff of the phonon effective theory. For $c_s$ close to one, this is parametrically the same as the momentum cutoff $\Lambda_{\rm UV}$ and we get
\be
\alpha \,  \Lambda_{\rm UV}^2 \sim {(1-c^2_s)^2} \sim (1-c_s)^2 < (1-c_s)\; , \qquad c_s \sim 1 \; ,
\ee
in agreement with our bound above. In the non-relativistic limit, with $c_s \ll 1$, we must remember that the gap $\Delta$ actually corresponds to the {\em mass} of a particle, times $c^2$, which is one in our units. For instance, for liquid helium, $\Delta$ is of order of the rest energy of a helium atom. We thus have that the momentum cutoff is $\Lambda_{\rm UV} \sim c_s \, \frac{\Delta}{c^2}$, in which case our $\alpha$ becomes
\be
\alpha \, \Lambda^2_{\rm UV} \sim 1 \; , \qquad c_s \ll 1 \; ,
\ee  
which is very far from saturating our bound above.

This analysis is a simple extension of that in terms of the group velocity of Sec.~\ref{group}.
Notice however that for an effective theory with {\em negative}  $\alpha$, from \eqref{Im omega one loop} we see that the ratio $\Im \, \omega/|\Im \, k|$ is maximized for imaginary $k$. The bound that we provided above is still correct (up to a factor of three), now for the absolute value of $\alpha$. But a negative $\alpha$ gives a negative contribution to the group velocity of our excitation, which means that if $c_s \le 1$, there is no bound on a negative $\alpha$ coming from imposing subluminality of the group velocity.
This is thus an explicit example that shows that our constraints on correlation functions are more powerful than imposing subluminality of group velocities.

\section{Solids and framids}\label{solids}

It is interesting to see how microcausality is obeyed in the EFT of solids. For simplicity, we consider only isotropic (that is, ``amorphous") solids---see e.g.~Ref.~\cite{Esposito:2020wsn} for a discussion about this point. The quadratic part of the long-distance effective theory for the phonons reads \cite{Nicolis:2015sra}
\be
S= \dfrac{1}{2} \int d^4x \left[ \dot{\vec{\pi}} \,^2 - (c_L^2 -c_T^2) (\vec \nabla \cdot \vec{\pi} )^2   - c_T^2 \left(\partial_i \pi_j \right)^2\right],
\ee
where $\vec{\pi}$ is a triplet of phonon fields transforming as a vector under rotations, and $c_L$ and $c_T$ are the propagation speeds for the longitudinal and transverse components of $\vec \pi = \vec \pi_L + \vec \pi_T$,
\be
\vec{\nabla} \times \vec{\pi}_L = 0 \; , \qquad \vec{\nabla} \cdot \vec{\pi}_T = 0 \; . 
\ee
Despite the different symmetry breaking pattern, the quadratic action for the EFT of the hypothetical ``framids" is exactly the same~\cite{Nicolis:2015sra}, but there the Goldstone fields are associated directly with the spontaneously broken boosts rather than with translations. At the level of two-point functions, we can thus treat the two cases simultaneously. 

The commutator Green's function $\C^{ij} = \langle \, [\pi^i, \pi^j] \, \rangle$ in the $(t,\vec{k})$ representation is 
\begin{equation}\C^{ij}(t, \vec k \, )= -i \;\hat{k}^i \hat{k}^j \, \dfrac{\sin (c_L k \, t)}{c_L k} -i  \left(\delta^{ij} -\hat{k}^i \hat{k}^j \right)\dfrac{\sin (c_T k \, t)}{c_T k} \; ,
\end{equation}
which has a manifest longitudinal-plus-transverse structure,

Notice that we are now dealing with the commutator of  fields  that are not scalars under rotations. So, we will have to check analyticity in the {complex vector} $\vec{k}$ rather in the scalar combination $k^2$. 
Clearly, the longitudinal and transverse parts of $\C^{ij}$ are {\it not} separately analytic in $\vec k$, since the unit vector $\hat{k} = \vec k / k$ is not analytic. But how can the non-analyticities cancel in their sum, given that the longitudinal and transverse contributions involve two independent parameters, $c_L$ and $c_T$? The situation becomes clearer if we reorganize the different structures thus:
\be
\C^{ij}(t, \vec k \, ) =-i \; \hat{k}^i \hat{k}^j \, \Big[\dfrac{\sin (c_L k \, t)}{c_L k} - \dfrac{\sin (c_T k \, t)}{c_T k} \big]
-i \;\delta^{ij} \, \dfrac{\sin (c_T k \, t)}{c_T k}  \; .
\ee
The $\delta^{ij}$ piece is analytic in $k^2$, and thus in $\vec k$. Now, however, it is clear how the other piece is analytic as well: when Taylor expanding the sines, the leading terms exactly cancel---regardless of the values of $c_L$ and $c_T$---and one is left with the structure
\be
\hat{k}^i \hat{k}^j \, \Big[\dfrac{\sin (c_L k \, t)}{c_L k} - \dfrac{\sin (c_T k \, t)}{c_T k} \big] = k^i k^j t^3 f(k^2 t^2) \; ,
\ee
where $f$ is an analytic function, and the offending unit vectors are gone.
This may sound surprising, but is in fact required by consistency with microcausality!
As to exponential boundedness, it works exactly like before: it is obeyed if and only if $c_L$ and $c_T$ are both subluminal.

A consequence of all this is that the fields $\vec \pi_L$ and $\vec \pi_T$ are not local fields, while their sum $\vec \pi$ is. Restricting one's attention to the longitudinal or transverse sector does not preserve locality, and thus microcausality. This is perhaps not surprising, since extracting $\vec \pi_L$ and $\vec \pi_T$ from $\vec \pi$ requires applying non-local projectors.

\paragraph*{Is analyticity always satisfied in EFT?}

The previous examples could give the impression that analyticity is always satisfied in a low energy EFT, perhaps after some magic cancellations. In fact this is not the case, as the following example demonstrates. Consider the quadratic action:
\begin{equation}
S^{(2)}= \dfrac{1}{2} \int dt d^3 x \left[ \alpha (\partial_i \partial_t \pi)^2 -\beta (\partial_i \partial^i  \pi)^2 \right] \; .
\end{equation}
The Feynman propagator is easily derived
\begin{equation}
\dfrac{1}{k^2} \dfrac{-i}{\alpha \omega^2- \beta k^2 + i\epsilon}\; ,
\end{equation}
and from this it is easy to see that the commutator does not have the required analyticity properties, since after isolating the real part (see App.~\ref{app:generalities}) and Fourier transforming in time we are left with a non-analytic factor of $1/k^2$.
As a consequence, there is no regime of validity where this EFT can be made consistent with microcausality. A scalar field with this quadratic action is sometimes known as a ``khronon'' and the fact that this theory has instantaneous propagation was already noticed in~\cite{Blas:2011ni,Creminelli:2012xb}.

A similar result can be obtained for generic high-derivative generalizations of solids of the form
\begin{equation}
S^{(2)}= \int \alpha (\vec{\nabla} \cdot  \partial_t \vec{\pi})^2 + \beta (\vec{\nabla} \times  \partial_t \vec{\pi})^2 + \gamma \vec{\nabla}^2 (\vec{\nabla} \cdot  \vec{\pi})^2 + \delta \vec{\nabla}^2(\vec{\nabla} \times  \vec{\pi})^2 \; .
\end{equation}

\section{Normal fluids}\label{fluids}

\subsection{Relativistic hydrodynamics} \label{hydro}

An interesting example to study is that of relativistic hydrodynamics (see also Ref.~\cite{Heller:2022ejw} for a complementary study of causality constraints in this context) . In a long-distance, low-frequency expansion, correlation functions in hydrodynamics can be derived purely from conservation laws and linear response theory. Restricting for simplicity to the correlation functions of the momentum density $T^{0i}(x)$, one finds the retarded Green's function \cite{Kovtun:2012rj}
\be
\tilde G^{ij}_R(\omega, \vec k \,) =i \, {k}^i {k}^j \, \bar w \, \frac{c_s^2 - i \omega \gamma_s}{\omega^2 - c_s^2 k^2 + i \gamma_s \, \omega k^2} + (\delta^{ij} k^2 - k^i k^j) \eta \frac{i}{i \omega - \gamma_\eta k^2}  \; ,
\ee
where $\bar w$ is the equilibrium enthalpy density, $\eta$ is the shear viscosity, and $\gamma_s$ and $\gamma_\eta$ are diffusion constants related to the shear and bulk viscosity. We recognize again a longitudinal-plus-transverse structure, with the former being associated with sound waves and the latter with the  shear modes. This correlation function is the correct one for so-called first-order hydrodynamics, which corresponds to the first non-trivial order in a derivative expansion beyond the perfect fluid limit.

Notice that now the longitudinal and transverse tensor structures involve directly the vector $\vec k$, in an analytic way, rather the unit vector $\hat k$. We can thus ignore them for our checks of analyticity. 

As a function of $\omega$, $\tilde G_R$ has two sound poles, at
\be \label{hydro poles 1}
\omega_{\rm sound}^\pm \simeq \pm c_s k - i \frac{\gamma_s}2  \, k^2 \; ,
\ee
and one shear pole, at
\be \label{hydro poles 2}
\omega_{\rm shear} = - i \gamma_\eta k^2 \; .
\ee
Notice that, consistently with the derivative expansion, we can only trust the approximate expression for $\omega_{\rm sound}^\pm$ up to $k^2$ order,  since higher order corrections would be affected by higher order terms in the derivative expansion, which we did not keep.
Performing the frequency integrals, after some straightforward algebra for the sound modes we find
\begin{align}
\frac{c_s^2 - i  \gamma_s \, \omega}{\omega^2 - c_s^2 k^2 + i \gamma_s \, \omega k^2} & \to 
- \theta(t) e^{-\frac{\gamma_s}{2}k^2 t} \times \\
&  \Big[ \frac{1}{1+\frac{\gamma_s^2}{4c_s^2}k^2} 
\Big( c_s \frac{\sin (c_s k \, t)}{k} - \frac{\gamma_s}{2} \cos (c_s k \, t)\Big) + \gamma_s \cos (c_s k \,  t) \Big] \; .
\end{align}
All pieces are analytic  in $k^2$, with the usual  subtlety about $\frac{\sin z}{ z}$, and, more importantly, with a subtlety about the pole at
\be
k^2  = - \frac{4 c_s^2}{\gamma_s^2} \; .
\ee
This is outside the regime of validity of the hydrodynamic derivative expansion, and  we should not push our analysis to such high values of $k^2$. To see this, it is enough to look at the expansion of the sound pole frequencies, \eqref{hydro poles 1}. When $k$ becomes of order $c_s/\gamma_s$, the expansion breaks down.

For the shear modes, we simply get
\be
\frac{1}{i \omega - \gamma_\eta k^2} \to - \theta(t) e^{- \gamma_\eta k^2 \, t} \; ,
\ee
which is clearly analytic in $k^2$.

As to  exponential boundedness, the retarded Green's function evolves in time according to the pole's complex frequencies \eqref{hydro poles 1}, \eqref{hydro poles 2}. In order to be consistent with microcausality, these must obey
\be
{\Im \, \omega}  < |{\Im \, k}| \; .
\ee
This is guaranteed within the regime of validity of the derivative expansion, provided of course $c_s$ is subluminal.
To see this, notice that for both sound and shear poles, the strongest constraints come from taking $k$ purely imaginary. We find, respectively,
\be
c_s + \frac{\gamma_s}{2} \Im \, k < 1 \; , \qquad  {\gamma_\eta} \,  \Im \, k < 1 \; .
\ee
The former constraint is tighter in general, because $c_s > 0 $ and $\gamma_s \gtrsim \gamma_\eta$ \cite{Kovtun:2012rj}, so for simplicity let us consider just that. Similarly to what we did for superfluids, we can interpret that constraint as a bound on the momentum cutoff of the theory:
\be \label{hydro cutoff}
 \Lambda_{\rm UV}  \lesssim \frac{1- c_s}{\gamma_s}
\ee 
As we argued above, we expect the derivative expansion to be break down at $k \sim c_s/\gamma_s$.
This cutoff scale obeys the parametric  bound \eqref{hydro cutoff}, unless the sound speed is very close to one. We do not know of any fluid with such high sound speeds---in fact, at high temperatures or densities fluids are generically expected to approach the scale invariant value $c_s^2 =1/(d-1)$, where $d$ is the number of spacetime dimensions (see however~\cite{Joyce:2022ydd,Nicolis:2023pye} for counterexamples). It would be interesting to sharpen these arguments and maybe show that the existence of such a fluid would create a tension with microcausality.

The example of relativistic hydrodyamics is instructive also because of recurrent claims in the literature that first-order hydrodynamics is inconsistent with relativistic causality. As emphasized by many---see e.g.~\cite{Baier:2008}---this is true only if one extrapolates the theory down to distance and time scales beyond the regime of validity of the derivative expansion, which one should not do. Our analysis is in perfect agreement with this conclusion.

\subsection{Gravity waves in the ocean}\label{ocean}
Another instructive example of our constraints is furnished by the so-called (surface) gravity waves in the ocean. Not to be confused with {\it gravitational} waves, these are simply the vertical-displacement waves that we see on the ocean's surface \cite{Landau6}. 

Consider an infinitely deep, incompressible ocean in the presence of a constant gravitational acceleration $g$. The dispersion relation for gravity waves is \cite{Landau6}
\be \label{gravity waves}
\omega_k = \sqrt{g \, k} \; , 
\ee
where $\vec k$ is the horizontal wavevector. Such a dispersion relation is highly non-analytic, which is perhaps not surprising, since in order to get it one has to integrate out the bulk degrees of freedom. Also, our constraints do not say anything directly about the analyticity of the dispersion relation, only about the analyticity of certain Green's functions. Still, the group velocity is
\be
v_g = \frac{d \omega_k}{d k} = \sqrt{\frac{g}{2 k}} \; ,
\ee
which becomes infinite in the $\vec k \to 0$ limit.

Of course the dispersion relation above is derived in the non-relativistic limit, where there is no light-cone, and so no microcausality to talk about. However, the only propagating  degrees of freedom in the bulk are the sound waves, which suggests that gravity waves should not propagate faster than the speed of sound. That is, they should obey {\it sonic} microcausality. Now, eq.~\eqref{gravity waves} assumes that the ocean is incompressible, which, in particular, implies that the speed of sound is infinite. In order to check sonic microcausality, we need to drop the incompressibility assumption.

Using $g$ and the speed of sound, $c_s$, we can build a typical wavenumber scale and a corresponding frequency scale:
\be
\kappa \equiv \frac{g}{2 c_s^2} \; , \qquad  \Omega \equiv c_s \kappa \; ,
\ee
where the factor of 2 is useful for what follows. It is natural to expect  the dispersion relation \eqref{gravity waves} to get modified as
\be
\omega_k = \sqrt{g \, k} \, f(k/ \kappa) \; ,
\ee
for some function $f$, such that sonic microcausality holds. For example, one could expect a smooth transition between \eqref{gravity waves} at high $k$, and a sound wave-like dispersion relation, $\omega_k = c_s k$, at low $k$. An $f$ accomplishing this is, for example, $f(x) = \tanh(\sqrt{x/2})$. In fact, the situation is much more involved, and it requires analyzing in detail the analytic structure of the retarded Green's function in frequency space. 

For an infinitely deep, \textit{compressible} ocean with, for simplicity, constant $c_s$, the surface-to-surface retarded Green's function for a suitably normalized vertical displacement field is\footnote{Computing such a Green's function involves integrating out the bulk degrees of freedom with suitable boundary conditions at infinite depth: the fields should go to zero or be purely outgoing waves there. These are the correct boundary conditions for the causal response of the ocean to sources localized at the free surface. Details of the computation will appear elsewhere.}
\be \label{GR ocean}
\hat G_R(\omega, \vec k \, ) = -2i \, \frac{2 k^2 - \omega^2 + \omega^2 \sqrt{1 + k^2 - (\omega+i \epsilon)^2}}{(\omega+i \epsilon)^4 - 4 k^2 } \; ,
\ee
where, to avoid clutter, we are measuring $\omega$ and $k$ in units of the $\Omega$ and $\kappa$ defined above. (In particular, $c_s = 1$ in these units.) The choice of the square root is such that the branch cut is on the negative real axis of its argument.

We are interested in
\be \label{ocean in time}
G_R(t, \vec k \, ) = \int \frac{d \omega}{2 \pi } \, \hat G_R(\omega, \vec k \, ) e^{-i \omega t } \; .
\ee
This is the object that should be analytic in $\vec k$ (or, for simplicity, in $k^2$), and exponentially bounded.
Consider for now $\vec k$ real, that is, $k^2$ real and positive. At first sight, \eqref{GR ocean} has four poles in the complex $\omega$ plane, at
\be \label{ocean poles}
\omega = \pm \sqrt{2 k} - i \epsilon \; , \qquad \omega = \pm i \sqrt{2 k} \; ,
\ee
and two cuts, starting at
\be
\omega = \pm \sqrt{1 + k^2}  \; ,
\ee
and extending, respectively, to $\pm \infty$.

The first pair of poles in \eqref{ocean poles} corresponds {\it exactly} to the dispersion law \eqref{gravity waves}---$g =2$ in the units that we chose. That is, there is {\it no} modification to the gravity waves' dispersion law due to compressibility. Were we to keep only the contribution of these poles, we would get
\be \label{poles with cs}
\tilde G_R(t, \vec k \, ) = \theta(t) \, \big( k -1 \big) \, \frac {\sin \big( \sqrt {2 k} \, t \big)}{\sqrt {2 k} } \; , 
\ee
which is, most definitely, {\it not} analytic in $k^2$ at $k^2=0$. Clearly, the other non-analyticities of \eqref{GR ocean} must play an important role, at least for small $k^2$.

\begin{figure}
\begin{center}
\includegraphics[width=14cm]{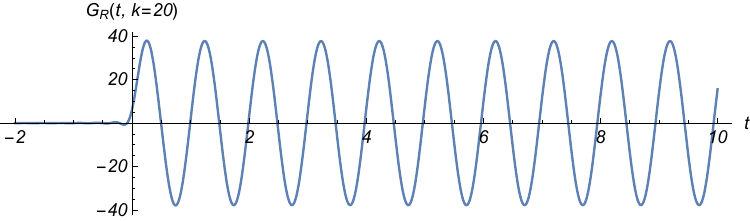}\\[.7cm]
\includegraphics[width=14cm]{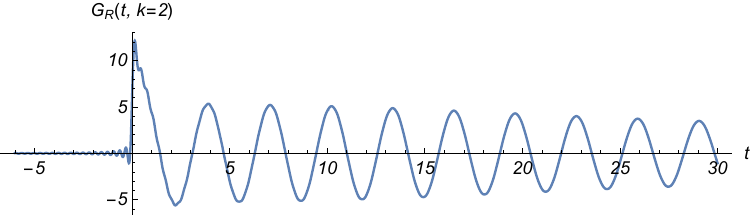}\\[.7cm]
\includegraphics[width=14cm]{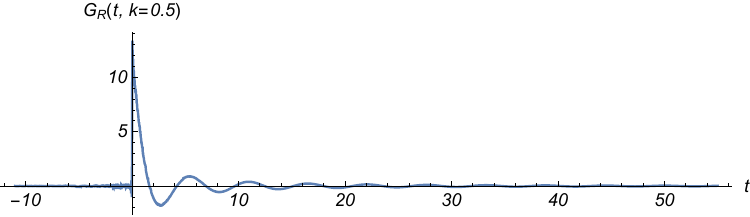}
\caption{\it \small \label{abyss} The time-dependence of the surface-to-surface retarded Green's function $\tilde G_R(t, \vec k)$ for gravity waves in an infinitely deep compressional ocean, from large horizontal wavenumbers (top) to small ones (bottom). At wavenumbers smaller than one (in the units discussed in the text), there are no waves propagating at the surface.}
\end{center}
\end{figure}

Let us thus analyze the analytic structure of \eqref{GR ocean} more carefully, keeping $k^2$ complex. It is not difficult to prove that, given our choice for the square root, the second pair of poles in \eqref{ocean poles} is always on the second sheet, and thus irrelevant for our purposes, whereas the first pair---the one giving rise to \eqref{poles with cs}---is on the first sheet if and only if
\be
|\Re \,k | = \big|\Re\,\sqrt{k^2} \big| > 1 \; .
\ee
In particular, those poles also move to the second sheet at small enough $\vec k$. And, for $\vec k$ going to zero, there are no poles on the first sheet, but only cuts. The cuts correspond to the bulk sound-wave continuum. This means that at low-enough $\vec k$, there are no surface waves! Any disturbance sourced at the surface quickly disappears into the depths. This is vividly demonstrated in fig.~\ref{abyss}.

\begin{figure}
\begin{center}
\includegraphics[width=15cm]{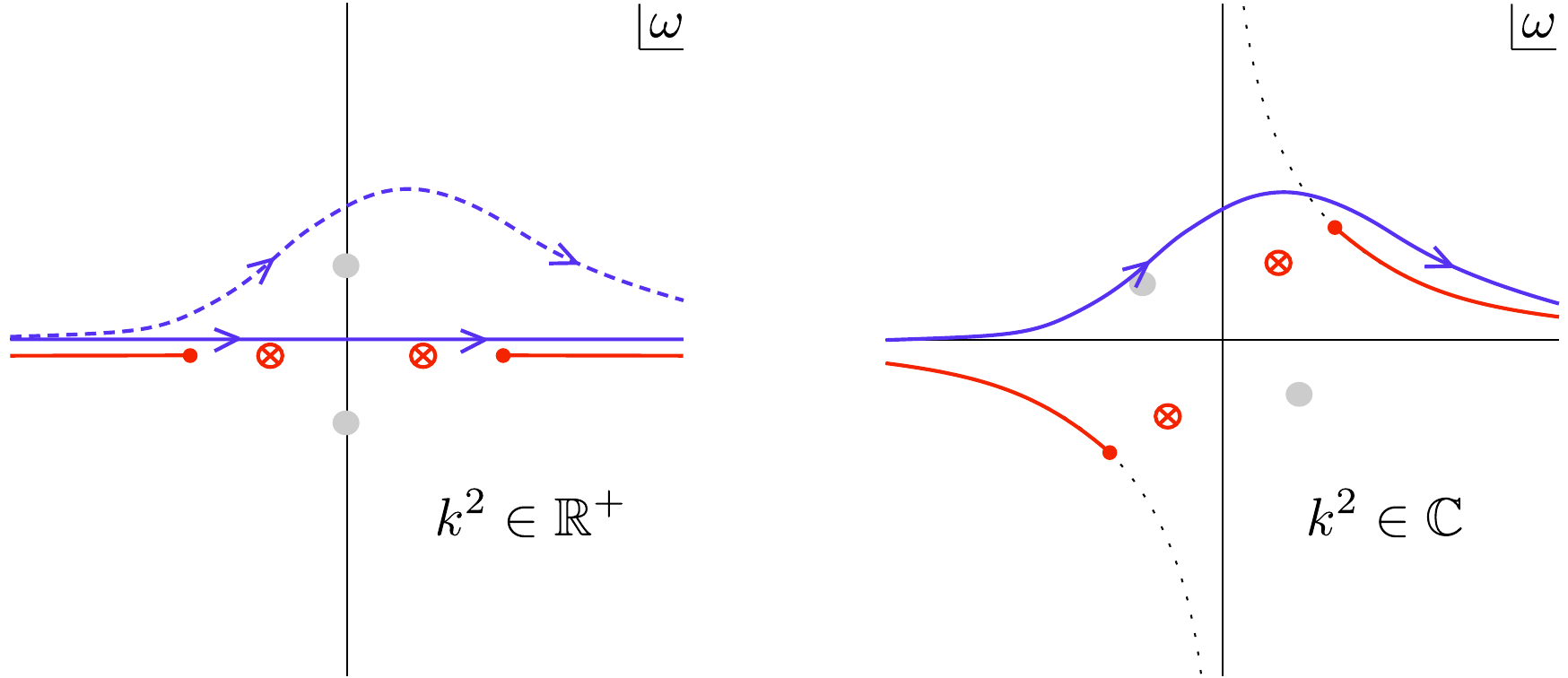}
\caption{\it \small \label{contour} The analytic structure and contours described in the sect.~\ref{ocean}. The greyed-out poles are in the second sheet, and one need not worry about them. The other poles also move to the second sheet for $|\Re \, k | < 1$.}
\end{center}
\end{figure}

As for the cuts, they are located at 
\be \label{ocean cuts}
\omega^2 = 1  + \mu^2 + k^2 \; , \qquad \mu^2 \in \mathbb{R}^+ \; ,
\ee
where $\mu^2$ can be thought of as a coordinate along the cuts. 
For fixed $\mu^2$, this equation is simply the dispersion relation for a relativistic massive particle, with squared mass $M^2 = 1+ \mu^2 > 0$. In such a case, as proved in Appendix \ref{imaginary}, for all complex $k$'s one has
\be
|\Im \, \omega | \le |\Im \, k| \; ,
\ee
with the equality being attained only for $\Im \, k = 0$. Moreover, splitting $\omega$ and $k^2$ into their real and imaginary parts, the imaginary part of \eqref{ocean cuts} yields
\be
\Re(\omega) \Im(\omega) = \sfrac{1}{2} \, \Im(k^2) \; .
\ee
This equation tells us that the cuts lie along hyperbolae in the complex $\omega$ plane, identified by the imaginary part of $k^2$. After originating at the branch points ($\mu^2 = 0$), the cuts approach the real axis as
\be
\Im \, \omega  \propto \frac{ 1}{ \Re \,\omega } \; .
\ee
The analytic structure just described is shown in fig.~\ref{contour}.

We thus see that, for fixed but arbitrary complex $k^2$, the non-analyticities of \eqref{GR ocean} extend only up to some finite value of $\Im(\omega)$, and approach the real axis at $\Re(\omega) \to \pm \infty$.  This allows us to {\it redefine} \eqref{ocean in time} for arbirtrary complex $k^2$ as the integral along a continuously deformed contour that avoids all the singularities but coincides with the original one for $\Re(\omega) \to \pm \infty$, such as that depicted in fig.~\ref{contour}.
Such a definition coincides with the original one in \eqref{ocean in time} for real, positive $k^2$, but is however analytic for all complex $k^2$, because it is a convergent integral of an analytic function of $k^2$, eq.~\eqref{GR ocean}. 
We thus reach the conclusion that, even though we are not able to compute the integral \eqref{ocean in time} explicitly, its output can be analytically extended to the whole complex $k^2$ plane.

We now come to the question of exponential boundedness. According to our definition of $G_R(t, \vec k \, )$ for complex $k^2$, we have
\be
\tilde G_R(t, \vec k \, )=  \int_\gamma \frac{d \omega}{2 \pi } \, \hat G_R(\omega, \vec k \, ) e^{-i\, \Re(\omega) t } e^{\Im(\omega) t} \; ,
\ee 
where $\gamma$ is the contour avoiding all singularities we just described. Up to non-exponential factors, we thus have
\be
| \tilde G_R(t, k^2) | \lesssim e^{\max_\gamma(\Im \, \omega) \, t} \; ,
\ee
and $\gamma$ can be pushed as low as the branch cuts and the poles on the first sheet. 

As noted above, on the cuts we have $|\Im \, \omega | \le |\Im \, k|$.
The poles on the first sheet obey
\be \label{imaginary part pole}
\omega_{\rm pole}^2 = 2 k
\ee
Taking the imaginary part, we have
\be
\Re(\omega_{\rm pole}) \Im(\omega_{\rm pole}) = \Im \,k \; .
\ee
However, recall that these poles are on the first sheet only for $|\Re(k)| \ge 1$, which, using \eqref{imaginary part pole}, implies
\be
\Re(\omega_{\rm pole})^2 \ge \Im(\omega_{\rm pole})^2 + 2 \ge 2 \; .
\ee
Putting these two equations together we find
\be
|\Im \,\omega_{\rm pole}| \le \frac{|\Im \, k|}{\sqrt{2}} < |\Im \, k|\; .
\ee

In conclusion, our retarded Green's function is analytic in $\vec k$ and bounded by
\be
| \tilde G_R(t, \vec k) | \lesssim e^{|\Im \, k| \, t} \; , 
\ee
and thus obeys sonic microcausality, as expected.

\section{Microcausality in cosmology}\label{cosmology}
 
We now switch to curved spacetime and study if and how microcausality can be used to derive bounds on effective field theories in cosmological backgrounds.\footnote{For an incomplete list of alternative approaches to positivity bounds in cosmological backgrounds see~\cite{Baumann:2019ghk,deRham:2021bll,Green:2023ids,CarrilloGonzalez:2023emp,Donath:2024utn} and references therein.} In this section, for convenience, we adopt the mostly plus signature.

\subsection{Scalar field in de Sitter space}

We start by analyzing a massless scalar field in de Sitter space.
We want to check how its two-point function satisfies the analyticity and boundedness implied by microcausality through the Paley-Wiener theorem.
We adopt flat slicing coordinates for de Sitter
\begin{equation}\label{eq:flatslice}
ds^2 = \dfrac{1}{H^2 \tau^2} \Big(- d\tau^2 +  d\vec{x} \,^2 \Big) ,
\end{equation}
where $H$ is constant and $-\infty<\tau <0$.
This is the de Sitter version of Poincar\'e coordinates.  In particular, the lightcone corresponds to $|\Delta \vec x| = |\Delta \tau|$.
Consider now a massless scalar in de Sitter,
\begin{equation}
S_{\rm free,dS}= \int \sqrt{-g} \left(- \dfrac{1}{2} g^{\mu\nu} \partial_\mu \phi \partial_\nu \phi \right)  d^4x= \frac{1}{2} \int \dfrac{1}{H^2 \tau^2} \left((\partial_\tau \phi)^2 - (\partial_i \phi)^2 \right)  d\tau d^3x \,.
\end{equation}
Choosing the Bunch-Davies vacuum by requiring that the two point function of $\phi$ reduces to the flat space result at early times with the correct normalization, we obtain the mode functions\footnote{The normalization can be determined by imposing that the mode functions be orthonormal with respect to the Klein-Gordon inner product.}
\begin{equation}
f_k (\tau) = \dfrac{H}{\sqrt{2k^3}} \left(1+ i k \tau \right)e^{-i k \tau },
\end{equation}
where we used the notation of appendix~\ref{app:generalities}.
To check microcausality we need to check analyticity and boundedness of 
\begin{align}
\C(\tau_1,\tau_2,k) & =  2i \, {\rm Im }\left( f_k(\tau_1) f_k^*(\tau_2) \right) \\
& = i \, \dfrac{H^2}{k^3} \Big( k(\tau_1-\tau_2) \cos\big(k(\tau_1-\tau_2)\big)-(1+k^2 \tau_1 \tau_2) \sin\big(k(\tau_1-\tau_2)\big) \Big) \; . \nonumber
\end{align}
It is straightforward to check that this is analytic in complex $k^2$.
Moreover, in the complex $k$ plane it is exponentially bounded by 
\begin{equation}\label{eq:bound_dS}
\vert \C(\tau_1,\tau_2,k) \vert \lesssim e^{|\Im \, k | \, |\tau_1-\tau_2|} \; .
\end{equation}

This result is consistent with microcausality in the fixed de Sitter background~\eqref{eq:flatslice}, whose lightcone is determined by the condition $- d\tau^2 +  d\vec{x} \,^2=0$. For fixed $\tau_1, \tau_2$, microcausality implies the vanishing of the position space commutator for $(\vec{x}_1-\vec{x}_2)^2 > (\tau_1 - \tau_2)^2 $, which is equivalent to~\eqref{eq:bound_dS} through the Paley-Wiener theorem.

\subsection{The EFT of inflation}

We now consider the case in which the universe undergoes a phase of near-exponential inflation driven by a scalar field $\phi$. The scalar evolves according to a homogeneous time-dependent background solution $\phi_0(t)$, while the metric is a flat FRW background
\begin{equation}
ds^2 = - dt^2 + a(t)^2 d\vec{x} \, ^2,
\end{equation}
with $H(t)\equiv \dot{a}/a$. It is also useful to define the slow-roll parameters
\begin{equation}
\epsilon \equiv - \dfrac{\dot{H}}{H^2}, \qquad \eta \equiv \dfrac{\dot{\epsilon}}{H \epsilon}.
\end{equation}
The dynamics of the scalar field and of gravity are now mixed and we cannot treat $\phi$ as a simple test field on a fixed background. 
We use comoving (unitary) gauge in which the perturbations are defined by the conditions
\begin{equation}
\phi(t,\vec{x})= \phi_0(t), \qquad h_{ij} = a^2(t)\, e^{2\zeta(t,\vec{x})} \, e^{ \gamma_{ij}} \;.
\end{equation}
$h_{ij}$ is the induced metric on spatial slices defined by $\phi = \rm const$, $\zeta$ denotes the curvature perturbation in comoving gauge, sometimes called $\mathcal{R}$, and $\gamma_{ij}$ is transverse and traceless and parametrizes tensor modes. 

The effective action for perturbations can be derived in great generality, without specifying the detailed dynamics of the underlying UV model, using the effective field theory of inflation~\cite{Cheung:2007st}.

At quadratic order, for curvature perturbations one gets
\begin{equation}\label{eq:zetaactionEFT}
S_{\zeta} = \frac{1}{2}\int d t d^3x \, \frac{2 M_{\rm Pl}^2 \epsilon}{c_s^2} a^3 \Big( \dot \zeta^2- c_s^2 \frac{(\partial_i\zeta)^2}{a^2} \Big),
\end{equation}
where $M_{\rm Pl}^2 \equiv 1/(8\pi G)$ and
the sound speed $c_s$ is defined by
\begin{equation}\label{eq:csEFT}
\dfrac{1}{c_s^2} = 1- \dfrac{2 M_2^4}{M_{\rm Pl}^2 \dot{H}} \; ,
\end{equation}
where $M_2(t)^4$ is the coefficient of the operator $(\delta g^{00})^2 / 2$ in the EFT of inflation. For instance, for a theory with action $S=\int \sqrt{-g}\, P(X,\phi)$, it is given by
\begin{equation}\label{eq:M2_PXphi}
M_2(t)^4 = \dot{\phi}^4_0(t) \dfrac{\partial^2 P(X,\phi)}{\partial X^2}\Bigg\vert_{\phi_0} , \qquad  {\rm where} \qquad X = g^{\mu\nu}\partial_\mu\phi \partial_\nu \phi.
\end{equation}
Higher derivative corrections give Hubble-suppressed terms, which are negligible as long as $H$ is much smaller than the scales $M_n$ appearing in the EFT. For a $P(X,\phi)$ theory this corresponds to requirement that the UV scale $\Lambda$ suppressing higher powers of $X$ be parametrically smaller than $M_{\rm Pl}$, with $\Lambda^2/ M_{\rm Pl}^2 \ll  \epsilon $.

It is convenient to adopt a new time variable $T$, which is related to conformal time but takes into account a possible time-dependence of $c_s$,
\be
dT \equiv \frac{c_s(t)}{a(t)} dt \; .
\ee
The quadratic action simply becomes
\be \label{zeta with Z}
S_{\zeta} = \frac{1}{2}\int d T d^3x \, Z(T) \Big( \zeta' \,^2- (\partial_i\zeta)^2 \Big) \; , 
\qquad 
Z(T) \equiv \frac{2 M_{\rm Pl}^2 \epsilon(T)}{c_s(T)} a^2(T)  \; ,
\ee
where primes denote derivatives w.r.t.~$T$.
The equation of motion for the mode functions takes the form
\begin{equation}
\zeta_k''+ \dfrac{Z'}{Z} \,\zeta_k' +  k^2 \, \zeta_k =0 \; ,
\end{equation}
where the ``Hubble friction" coefficient is
\begin{equation}\label{eq:csmode}
 \dfrac{Z'}{Z}= \frac{aH}{c_s}\left(2+ {\eta-s}\right), \qquad {\rm with} \quad s\equiv \dfrac{\dot{c}_s}{H c_s}  \; .
\end{equation}

Up to now we did not make any slow-roll approximation. In order to find explicit solutions to this equations and derive constraints from microcausality we will now assume the slow-roll conditions that $\epsilon, \eta, s$ are much smaller than one, and solve our equation by keeping first-order terms only,\footnote{To all orders, the function $a H/c_s$ obeys the differential equation $(a H/c_s)' = (a H/c_s)^2 \cdot (1-\epsilon-s)$. The solution to first order in slow roll is $a H/c_s \simeq - (1+\epsilon+s)/T$.}
\begin{equation}
 \zeta_k''+ \dfrac{1-2 \nu}{T} \,\zeta_k' +  k^2 \, \zeta_k =0, \qquad {\rm where} \quad \nu \equiv \dfrac{1}{2}\big(3 + 2 \epsilon +{\eta+s} \big) \; ,
 \end{equation}
where since the only explicitly time-dependent parameters are of first-order in the slow-roll expansion, we can consistently treat $\nu$ as constant.

The above equation can be solved in terms of Hankel functions,
\begin{equation}\label{eq:expansionInflation}
\zeta_k(\tau) = c_1 \cdot (-T)^\nu H^{(1)}_\nu(-kT) + c_2 \cdot {(-T)^\nu} H^{(2)}_\nu(-kT) \; .
\end{equation}
The solution that has positive frequency at early times, 
\begin{equation}
\zeta_k(T) \propto  e^{-i k \, T}, \qquad T \to -\infty
\end{equation}
 can be  singled out by considering the asymptotic expansion of the Hankel functions for $T \to -\infty$,
\begin{equation}
H^{(1,2)}_\nu( - k T)  \simeq \sqrt{\dfrac{2}{\pi (-kT)}} e^{\mp i(k T + \frac{\nu \pi}{2} + \frac{\pi}{4})}\; ,
\end{equation}
which tells us that $c_2$ should vanish.
The coefficient $c_1$ can be determined from the Klein-Gordon product normalization condition (see eq.~\eqref{eq:KGnorm} in the appendix). Keeping into account the additional non-canonical $Z$ factor  in the action~\eqref{zeta with Z}, after some simple algebra we find
\begin{align}
\zeta_k(\tau) & \simeq \dfrac{\sqrt{\pi}}{2 M_{\rm Pl}} \sqrt{\frac{-T}{Z(T)}} H^{(1)}_\nu(-k T) \\ 
& \simeq \dfrac{\sqrt{\pi}}{2 M_{\rm Pl}} \sqrt{\frac{(-T_\star)^{-2\epsilon-\eta-s}}{\epsilon_\star {c_s}_\star} } H_\star \cdot (-T)^\nu H^{(1)}_\nu(-k T) \; ,
\end{align}
where we chose an overall phase, $T_\star$ is an arbitrary time, and all starred quantities are evaluated there.

We consider now the constraints from microcausality. From eq.~\eqref{eq:commutatorK}, the  $\langle [\zeta, \zeta] \rangle$ commutator is
\begin{equation}
\C(T_1, T_2; k) = 2i \, {\rm Im }\left( \zeta_k(T_1) \zeta_k^*(T_2) \right) 
\propto \Im \left( H^{(1)}_\nu(-k T_1 )  H^{(1)}_\nu(-k T_2) ^* \right) 
\; ,
\end{equation}
and this has to be an analytic function of $k^2$ and exponentially bounded.
Let us first consider analyticity. 
It is convenient to express $H^{(1)}_\nu$ in terms of Bessel-$J$ functions: 
\begin{equation}
H^{(1)}_\nu(y) = \dfrac{J_{-\nu}(y) -J_{\nu}(y)e^{-i\nu \pi}}{i \sin(\nu \pi)}, \qquad H^{(1)}_\nu(y){}^* = \dfrac{J_{-\nu}(y) -J_{\nu}(y)e^{i\nu \pi}}{-i \sin(\nu \pi)},
\end{equation}
where we kept the argument $y$ real in taking the complex conjugate.
Then, up to irrelevant factors, our commutator is
\be
\C(T_1, T_2; k) \propto \Big(J_{-\nu}(-k T_2)J_{\nu}(-k T_1)- J_{-\nu}(-k T_1)J_{\nu}(-k T_2) \Big)
\ee
It so happens that the Bessel-$J$ functions can be written as
\be
J_{\nu}(y) = y^\nu f_\nu(y^2) \; ,
\ee
where $f_\nu$ is an entire function of complex $y^2$. We thus see that in the commutator above, for arbitrary $T_1$ and $T_2$, all non-integer powers of $k$ cancel, and we are left with an entire function of $k^2$, as required by microcausality.

To check exponential boundedness without having to control non-exponential factors, we can consider the case in which at least one of the two operators is inserted at a large negative time, $kT_{1} \to - \infty$ (with fixed $k$). 
We stress that in doing so we are not restricting ourselves to probing the dynamics only on subhorizon scales or, equivalently, to considering the local dynamics of small wavepackets in a weakly curved background. In fact, by studying the  $k T_1 \to -\infty$, $k T_2 \to 0$ case we are following a mode that gets stretched from subhorizon to superhorizon scales in the curved background. Using the asymptotic expansions of $H^{(1)}_\nu(y)$,
\begin{align}
&H^{(1)}_\nu(y) \simeq \sqrt{\dfrac{2}{\pi y}} e^{i(y - \frac{\nu \pi}{2} - \frac{\pi}{4})} ,\qquad {\rm for} \qquad y \to \infty \\
&H^{(1)}_\nu(y) \simeq -\dfrac{i \, \Gamma(\nu)}{\pi} \left(\dfrac{y}{2}\right)^{-\nu} ,\qquad {\rm for} \qquad y \to 0 
\end{align}
for $k T_1 \to -\infty$, $k T_2 \to 0$ we get that the commutator is of order
\begin{equation}
\C(T_1 , T_2 ; k)  \sim   \cos\left(-k T_1 - \frac{\nu \pi}{2} - \frac{\pi}{4}\right) \;,
\end{equation}
where we are omitting power-law prefactors.
For complex $k$ we thus have
\be
| \C(T_1 , T_2 ; k) | \sim e^{|\Im k| \, | T_1 |} = e^{|\Im k| \cdot | \int^{\tau_1} c_s(\tau) d \tau | } \; 
\ee
where $\tau$ is conformal time: similarly to the de Sitter case, the lightcone of an FRW metric in comoving coordinates is determined by the condition $- d\tau^2 +  d\vec{x} \, ^2=0$, so that microcausality implies the vanishing of the position space commutator for $(\vec{x}_1-\vec{x}_2)^2 > (\tau_1 - \tau_2)^2 $. We thus expect exponential boundedness for $\C$ in the form
\be
| \C(T_1 , T_2 ; k) | \lesssim e^{|\Im k| \cdot | \tau_1 - \tau_2 |} \; ,
\ee
which in our case is obeyed as long as $c_s(\tau) \leq 1$, not surprisingly.

We can repeat the same analysis for the case in which both operators are inserted at early times, $k T_{1,2}  \to -\infty $. We find
\begin{equation}
\vert\C(T_1, T_2; k) \vert \lesssim   e^{|\Im k| \cdot | \int^{\tau_2}_{\tau_1} c_s(\tau) d \tau | },
\end{equation}
which again obeys microcausality as long as $ c_s(\tau) \leq 1$. 

From eq.~\eqref{eq:csEFT}, we thus see that in a theory that satisfies the Null Energy Condition (which requires $\dot{H}\leq 0$), microcausality implies
\begin{align}
&M_2(t)^4 \geq 0 \quad {\rm in\, the \, EFT \, of \, Inflation,} \\
&\dfrac{\partial^2 P(X,\phi)}{\partial X^2}\Bigg\vert_{\phi_0(t)}\geq 0 \quad {\rm in\; a}\; P(X,\phi)\; {\rm theory} .
\end{align}
This bound is analogous to the similar statement derived in flat space. Notice however that in flat space one can argue that, in an interacting theory, the coefficient has to be strictly positive, whereas our argument only allows us to conclude that it must be non-negative. Moreover, the positivity statement relies on the fact that $\dot{H}\neq 0$ and does not directly apply to de Sitter space.

\subsection{The EFT for inflation}\label{EFTforInflation}

We can consider analogous bounds in the so-called effective field theory {\it for}  inflation~\cite{Weinberg:2008hq}, which corresponds to a different power-counting than that in the effective field theory {\it of} inflation~\cite{Cheung:2007st}.
Up to field redefinitions and total derivatives the effective action for the inflaton field and gravity up to four derivatives takes the form~\cite{Weinberg:2008hq}
\begin{align}
S_{\rm EH} = \int d^4x \sqrt{-g} \left( \dfrac{M_{\rm Pl}^2}{2} R - \dfrac{X}{2}  - V(\phi) + f_1(\phi)\dfrac{X^2}{\Lambda^4} + f_9(\phi) W^{\mu\nu\rho\sigma} W_{\mu\nu\rho\sigma}\right) ,
\end{align}
where $X = g^{\mu\nu}\partial_\mu\phi \partial_\nu \phi$, $W^{\mu\nu\rho\sigma}$ is the Weyl tensor, and we dropped a parity odd $W \tilde{W}$ term that does not contribute to the quadratic action for scalar perturbations. The $f_1$ term contributes a correction to the sound speed of scalar perturbations in agreement with eqs.~\eqref{eq:csEFT} and \eqref{eq:M2_PXphi}, where 
\begin{equation}
\dfrac{\partial^2 P(X,\phi)}{\partial X^2}\Bigg\vert_{\phi_0} =  \dfrac{2f_1(\phi)}{\Lambda^4} .
\end{equation}
The $f_9$ term contributes a correction equivalent to $\Delta f_1(\phi) = f_9(\phi)\cdot(\Lambda^4/3M_{\rm Pl}^4)$~\cite{Weinberg:2008hq}. We are left with 
\begin{equation}
\dfrac{1}{c_s^2} = 1- \dfrac{4 \dot{\phi}_0(t)^4}{M_{\rm Pl}^2 \dot{H}} \left( \dfrac{f_1(\phi)}{\Lambda^4} + \dfrac{f_9(\phi)}{3M_{\rm Pl}^4} \right).
\end{equation}

The analysis of our previous section would then tell us that microcausality implies the positivity bound 
\begin{equation}
\left( \dfrac{f_1(\phi)}{\Lambda^4} + \dfrac{f_9(\phi)}{3M_{\rm Pl}^4} \right) \geq 0.
\end{equation}
While this conclusion is robust in the case in which the UV cutoff that controls the inflaton's self-interactions is $\Lambda \ll  M_{\rm Pl}$, a note of caution is warranted. In deriving bounds from microcausality we assumed a fixed gravitational background and used its light-cone to define microcausality. While this requirement seems reasonable in a framework in which gravity is treated as a classical background~\cite{Dubovsky:2007ac}, the notion of a lightcone become less well defined when dynamical gravity and field redefinitions for the metric are taken into account~\cite{Adams:2006sv, Bordin:2017hal}.  One possibility would be to define an underlying Minkowski space
and treat the metric just as a tensor field living in such a flat space \cite{Visser:1998ua, Adams:2006sv}, but this procedure appears to be plagued by gauge ambiguities~\cite{Gao:2000ga}. So, our best hope to take advantage of microcausality in curved space seems to be that of sticking with a fixed background. The effects of dynamical gravity, together with quantum gravitational corrections, pollute the sharpness of the lightcone with $M_{\rm Pl}$ suppressed corrections, so that our statements are informative only when  $M_{\rm Pl}$ suppressed effects can be neglected. This state of affairs closely parallels the case of S-matrix positivity bounds, where in the presence of dynamical gravity a $t$-channel pole in the forward limit complicates the analysis of gravitational theories, and in fact allows small regions of negativity~\cite{Caron-Huot:2021rmr}.

\section{Discussion}

We appear to live in a universe where Lorentz symmetry, if it's broken at all, is broken spontaneously. Microcausality is thus expected to be satisfied, regardless of what Lorentz-breaking state one happens to be considering.
As such, the commutator Green's function $G_c$ \eqref{Gcdef}, expressed as a function of time $t$ in (spatial) Fourier space,
should (1) be analytic in the complex vector $\vec k$, and (2) be exponentially bounded, according to the Paley-Wiener theorem \eqref{exp bound vec k}. These two statements combined can be shown to be equivalent to the vanishing of $G_c$ outside the lightcone in real space. When rotation is not broken, the analyticity property translates into analyticity in
the complex $k^2 \equiv \vec k \cdot \vec k$. And when considering an EFT, taking the
large $k t$ limit while staying within the regime of validity of the EFT, it's sufficient to focus on the exponential
in exponential boundedness \eqref{ktlarge}.

We consider the implications of these statements in a variety of settings,
from superfluids, solids/framids, normal fluids to inflation. A few recurring themes emerge from these examples:
(1) Not surprisingly, microcausality typically implies subluminal sound speed, though see Sec. \ref{group} on the possibility of superluminal group velocity. (2) In verifying microcausality in an EFT, it's crucial that one stays within the regime of validity of the EFT, i.e. $k$ and $\omega$ must stay below the cut-off of the theory in question. (3) In systems where multiple gapless degrees of freedom are present, analyticity sometimes requires cancelation among their respective contributions to the Green's function. An example is the solid/framid with its longitudinal and transverse modes (Sec. \ref{solids}). (4) In verifying the analyticity of the Green's function in $(t, \vec k)$ space, one might encounter an integral over $\omega$ for which care must be taken in choosing the integration contour. An example on gravity waves is discussed in Sec. \ref{ocean}. (5) Microcausality as stated so far concerns the two-point function, but in some cases, can be used to make statement about interactions. Examples can be found in the discussions on superfluid (Sec. \ref{simpleExamples}) and on inflation (Sec. \ref{EFTforInflation}), where one starts from a manifestly Lorentz invariant theory and proceeds to consider a Lorentz violating background.

The last point raises a natural question: can we formulate a statement of microcausality that can be applied directly to higher-point functions? Consider the following nested commutator:
\begin{equation}\label{nested3}
[{\cal O}_3 (t, \vec x_3),   [{\cal O}_2(t, \vec x_2), {\cal O}_1(0, 0)] ] \, .
\end{equation}
It's obvious that this vanishes if $(t, \vec x_2)$ and $(0, 0)$ are space-like separated.
Making use of the Jacobi identity, and the fact that ${\cal O}_3$ and ${\cal O}_2$ are at equal time and therefore commute, we see that this nested commutator also vanishes if $(t, \vec x_3)$ and $(0, 0)$ are space-like separated. Thus, the corresponding 3-point function:
\begin{equation}
G_{c}^{(3)}(t, \vec x_2, \vec x_3) \equiv \langle [{\cal O}_3 (t, \vec x_3),   [{\cal O}_2(t, \vec x_2), {\cal O}_1(0, 0)] ] \rangle 
\end{equation}
vanishes if $|\vec x_2|^2 > t^2$ or $|\vec x_3|^2 > t^2$. Fourier transforming into
$G_{c}^{(3)}(t, \vec k_2, \vec k_3)$ and invoking the Paley-Weiner theorem, we conclude
$G_{c}^{(3)} (t, \vec k_2, \vec k_3)$ is analytic in complex $\vec k_2$ and $\vec k_3$, and is bounded by
$e^{(\norm{{\rm Im\,} \vec k_3} + \norm{{\rm Im\,} \vec k_2}) \cdot |t|}$ up to prefactor.
This statement can be straightforwardly generalized to
\begin{equation}
  G_{c}^{(n)}(t, \vec x_2, ... \vec x_{n-1}, \vec x_n) \equiv \langle [{\cal O}_n(t, \vec x_n),  [{\cal O}_{n-1} (t, \vec x_{n-1}), .... [{\cal O}_2 (t, \vec x_2), {\cal O}_1 (0, 0)]...]] \rangle \, ,
\end{equation}
which vanishes if any of the $|\vec x_2|^2 , ..., |\vec x_n|^2$ exceeds $t^2$, resulting in analogous analyticity and boundedness statements on $G_{c}^{(n)}(t, \vec k_2, ... , \vec k_n)$. It would be interesting to use these higher-point Green's functions to constrain interacting theories with broken Lorentz, such as those that occur naturally in cosmology. We hope to explore this in the near future.

\section*{Acknowledgements}

We thank Paolo Creminelli, Riccardo Rattazzi, Borna Salehian and Franco Strocchi for discussions and comments. We are grateful to Lorenzo Gavassino, Federico Piazza, and especially Paolo Creminelli and Borna Salehian for comments on a preliminary version of this work. Our work was partially supported by the US Department of Energy (award no.~DE-SC0011941).

\appendix
\section*{Appendix}

\section{Some useful expressions for commutator two-point functions}
\label{app:generalities}
We collect here some results on commutators of scalar fields in homogeneous and isotropic backgrounds. This assumption is not essential but simplifies some of the computations, while being general enough to treat interesting cases such as cosmological backgrounds, including inflation, superfluids, fluids, and relativistic hydrodynamics.

\paragraph*{Free scalars.}
Consider a (real) free scalar field $\phi(t,\vec{x})$ and assume that in canonical quantization it is quantized in terms of classical mode functions $u_{\bf k}$ of the form 
\begin{equation}\label{eq:mf}
u_{\bf k}(t,\vec{x}) = f_k(t) e^{ i \vec{k}\cdot \vec{x}}, \qquad u^*_{\bf k}(t,\vec{x})  = f^*_k(t) e^{- i \vec{k}\cdot \vec{x}}.
\end{equation}
The field operator $\hat \phi$ and its conjugate momentum~$\hat{\Pi}$ are then given by
\begin{equation}
\hat \phi(t,\vec{x}) = \int \dfrac{d^3 k}{(2\pi)^3} \left(u_{\bf k}(t,\vec{x}) \hat{a}_{\bf k} + u^*_{\bf k}(t,\vec{x}) \hat{a}^\dagger_{\bf k} \right), \qquad  \hat{\Pi}(t,\vec{x}) = \sqrt{h}\, \partial_t \phi \, ,\end{equation}
where we denote by $h$ the spatial metric.
The mode functions are normalized so that 
\begin{equation}
\langle u_{\bf k} , u_{\bf k'}\rangle = (2\pi)^3 \delta^{(3)}({\bf k} - {\bf k'}), \qquad \langle u^*_{\bf k} , u^*_{\bf k'}\rangle = -(2\pi)^3 \delta^{(3)}({\bf k} - {\bf k'}),\qquad  \langle u_{\bf k} , u^*_{\bf k'}\rangle = 0,
\end{equation}
with respect to the Klein-Gordon inner product, see for instance~\cite{Birrell:1982ix}.
The normalization condition on the mode functions~\eqref{eq:mf} can be expressed as
\begin{equation}\label{eq:KGnorm}
i \sqrt{h}\sqrt{g^{00}} (f_k^* \partial_t f_k -\partial_t f_k^*  f_k) =1.
\end{equation}
With this normalization the equal-time canonical commutation relations 
\begin{equation}
[\hat \phi(t,{\bf x}),\hat{\Pi}(t,{\bf x'})]= i \delta^{(3)}({\bf x} - {\bf x'})
\end{equation}
are recovered for $[\hat{a}_{\bf k},\hat{a}^\dagger_{\bf k'}]=(2\pi)^3 \delta^{(3)}({\bf k} - {\bf k'})$, and $[\hat{a}_{\bf k},\hat{a}_{\bf k'}]=0$.

The commutator between two fields inserted at different spacetime points can now be easily computed and is given by
\begin{equation}
[\hat{\phi}(t_1,\vec{x}_1), \hat{\phi}(t_2,\vec{x}_2)] = \int \dfrac{d^3 k}{(2\pi)^2} \, 2i \, {\rm Im }\left( f_k(t_1) f_k^*(t_2) \right) e^{i \vec{k} \cdot (\vec{x}_1 - \vec{x}_2)}.
\end{equation}
Thanks to translational and rotational invariance, this is a function of $(\vec{x}_1 - \vec{x}_2)^2$ only and its Fourier transform with respect to spatial coordinates depends on the spatial momentum only through $k^2$. This expressions is valid as an operator statement, where on the right the identity operator is understood. It can be evaluated in arbitrary states or inserted in a generic multipoint correlation functions. When evaluated on the ground state it corresponds to the commutator Green's function: we will only focus on this correlator and leave an analysis of more general correlation functions to future work.

It is the useful to define the Fourier transform with respect to $(\vec{x}_1 - \vec{x}_2)$:
\begin{equation}\label{eq:commutatorK}
\C(t_1,t_2,\vec k) \equiv 2i \, {\rm Im }\left( f_k(t_1) f_k^*(t_2) \right).
\end{equation}
Since the commutator vanishes for space-like separations, $\C(t_1,t_2,k)$ is constrained by the Paley-Wiener theorem to be analytic in $k^2$ and exponentially bounded as reviewed in section~\ref{general}. We stress that this property is a necessary and sufficient condition for microcausality. 
\paragraph*{Commutator in terms of $T$-ordered two-point function.} When including interactions it is often convenient to study time-ordered correlators, and take advantage of Feynman rules and perturbative tools to compute loop correction. It is therefore useful to write the commutator Green's function in terms of the time-ordered two-point function. It is convenient to do this either in frequency or in position space. We find
\begin{align}
&\hat{G}_c(\omega,\vec{k}) = 2 \, {\rm sign}(\omega)\, {\rm Re}\, \hat{G}_T(\omega,\vec{k})\\
&G_c(t,\vec{x}) = 2i \, {\rm sign}(t)\, {\rm Im}\, {G}_T(t,\vec{x})
\end{align}
These relations are fully non-perturbative.
The former follows from a spectral representation of two-point functions---see e.g.~\cite{Weinberg:1995mt}. The latter follows immediately from the definitions of commutators and time-order correlators.

\section{Complex $\vec k$ vs.~complex $k$}\label{k vs k}
For any complex vector $\vec k \in \mathbb{C}^3$, define its real and imaginary parts
\be
\vec k = \vec R + i \vec I \; , \qquad \vec R, \vec I \in \mathbb{R}^3 \; .
\ee
Then,
\be
k^2 \equiv \vec k \cdot \vec k = R^2 - I^2 + 2 i \, R I \cos \theta \; ,
\ee
where $R \ge 0$ and $I \ge 0$ are the magnitudes of the corresponding real vectors, and $\theta$ is the angle between such vectors.

On the other hand, defining the complex number
\be
k \equiv \sqrt{k^2}
\ee
and splitting it into its real and imaginary parts,
\be
k = {\cal R} + i {\cal I} \; , \qquad {\cal R},{\cal I} \in \mathbb{R}\; ,
\ee
we have
\be
k^2 = k \, k =  {\cal R}^2 - {\cal I}^2 + 2 i \, {\cal R} {\cal I}  \; .
\ee
Equating the two expression for $k^2$ and squaring the imaginary parts we get
\begin{align} 
R^2 - I^2 & = {\cal R}^2 - {\cal I}^2  \label{RI1}\\
R^2 I^2 \cos^2 \theta & = {\cal R}^2 {\cal I}^2 \label{RI2} \; .
\end{align}

\begin{figure}
\begin{center}
\includegraphics[width=6cm]{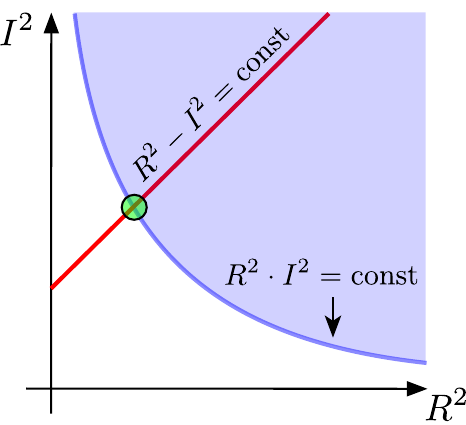}
\caption{\label{R2I2} \it\small The geometry of certain real and imaginary parts described in the text.}
\end{center}
\end{figure}

Now, if we have an analytic function of $k^2$ bounded by
\be
e^{||\Im \, \vec k\, || \cdot |t|} = e^{I \, |t|} \; ,
\ee
it must be in fact bounded by the minimum of such an exponential over all possible complex vectors $\vec k$ that have the same $k^2$, or, equivalently, the same $k = {\cal R} + i {\cal I}$.  Plotting the conditions \eqref{RI1}, \eqref{RI2} in the $(R^2,I^2)$ plane for fixed ${\cal R}$  and ${\cal I}$---see fig.~\ref{R2I2}---, and keeping into account that $\cos^2 \theta < 1$, it is clear that the minimum of $I$ for constant $k$ is achieved for $R^2 = {\cal R}^2$,  $I^2 = {\cal I}^2$, $\cos^2 \theta =1$.\footnote{Equivalently, it is achieved  for $\vec k = \hat n \, k$, where $\hat n$ is any real unit vector. The sign ambiguity in $\cos \theta = \pm 1$ is resolved by {\it not} squaring the imaginary part of $k^2$ as we did above.}

Therefore, our analytic function of $k^2$ should in fact be bounded by 
\be
e^{|{\cal I} | \, |t|} = e^{|\Im \, k|  \, |t|} \; .
\ee


\section{Analyticity in $(\omega, \vec k\,)$ space} \label{frequency space}
In the case of correlation functions in the mixed $(t, \vec k\,)$ representation, the analyticity and boundedness criterion is obtained for fixed $t$ or, in the distributional case, after smearing on time-dependent test functions of compact support. After Fourier transforming to frequency space, the analyticity statement one should check becomes: $\hat G_c(\omega, \vec k)$ is a $\vec k$-dependent distribution in $\omega$ such that, when applied to any {\it analytic} test function $f(\omega)$, yields an analytic function of $\vec k$. This is because compact support functions are mapped to analytic functions under Fourier transform.

Let us look at the example of a free scalar in a Lorentz invariant theory. The commutator two-point function in frequency space is 
\begin{equation}
\hat G_c(\omega, \vec k) = 2\pi \, {\rm sign}(\omega)\,  \delta(p^2-M^2).
\end{equation}
If we consider this as a distribution in $\omega$ on the space of analytic test functions $f(\omega)$, after integrating we obtain 
\begin{equation}
\int \hat G_c(\omega, \vec k) \, f(\omega) \dfrac{d\omega}{2\pi} = \dfrac{f(\omega_k)}{2\omega_k} - \dfrac{f(-\omega_k)}{2\omega_k} = \dfrac{f(\omega_k)-f(-\omega_k)}{2\omega_k} .
\end{equation}
It is easy to see that this is an even analytic function of $\omega_k=\sqrt{k^2+M^2}$, and thus an analytic function of $\vec k$.

The boundedness criterion is obtained after taking into account the exponential boundedness of the analytic test functions to which the compact support time-dependent test function are mapped under Fourier transform.

\section{$\Im  \, \omega_k$ vs.~$\Im \, k$}\label{imaginary}
We want to show that 
\be
|\Im(\omega_k)| \equiv |\Im(\sqrt{k^2+M^2}) | \le  | \Im(k)| \; , \qquad (M^2 \geq0, \; k\in \mathbb{C} ) \; .
\ee
This is trivially true for $M=0$, therefore Let us focus on the case $M\neq 0$.
Upon rescaling $k$ by $M$, this is the same as
\be \label{thesis}
|\Im(\sqrt{1+k^2}) | \le  | \Im(k)| \; .
\ee

To prove this, we split $k$ and $\sqrt{1+k^2}$ into their real and imaginary parts
\be
k = {\cal R} + i \, {\cal I } \; ,  \qquad \sqrt{1+k^2} = R +i \, I\; ,
\ee
and use the identity
\be
(\sqrt{1+k^2})^2 = 1 + (k)^2  \; ,
\ee
to obtain the equations
\begin{align}
R^2 -I^2 & = 1+ {\cal R}^2 - {\cal I}^2,\\
R^2 \cdot I^2 & = {\cal R}^2 \cdot {\cal I}^2.
\end{align}
The relevant geometry is still that of fig.~\ref{R2I2}. 

We see that $R^2$ and $I^2$ are two nonnegative numbers whose product is the same as that of two other nonnegative numbers, ${\cal R}^2$ and ${\cal I}^2$, but whose difference is larger. We must have
\be
R^2 \geq {\cal R}^2 \; , \qquad I^2 \le {\cal I}^2 \; ,
\ee
with the second equality being attained only for $I = {\cal I} = 0$.

Therefore, in the whole $k$ complex plane, we have
\be
|\Re\,  \omega_k| > |\Re \, k| \; , \qquad |\Im\,  \omega_k| \le |\Im \, k| \qquad\qquad (M^2 > 0 ) \; .
\ee

Notice that  the same analysis applies in the case of a tachyon, $M^2<0$, and yields the opposite inequalities,
\be
|\Re\,  \omega_k| < |\Re \, k| \; , \qquad |\Im\,  \omega_k| \ge |\Im \, k|  \qquad\qquad (M^2 < 0 ) \; .
\ee

\bibliographystyle{JHEP}
\bibliography{biblio}

\providecommand{\href}[2]{#2}\begingroup\raggedright\begin{thebibliography}{10}

\bibitem{Strocchi}
F.~Strocchi, \emph{Selected Topics on the General Properties of Quantum Field
  Theory}. World Scientific, 1993,
  \href{https://doi.org/10.1142/1807}{10.1142/1807}.

\bibitem{Dubovsky:2007ac}
S.~Dubovsky, A.~Nicolis, E.~Trincherini and G.~Villadoro, \emph{{Microcausality
  in curved space-time}},
  \href{https://doi.org/10.1103/PhysRevD.77.084016}{\emph{Phys. Rev. D}
  {\bfseries 77} (2008) 084016}
  [\href{https://arxiv.org/abs/0709.1483}{{\ttfamily 0709.1483}}].

\bibitem{Creminelli:2022onn}
P.~Creminelli, O.~Janssen and L.~Senatore, \emph{{Positivity bounds on
  effective field theories with spontaneously broken Lorentz invariance}},
  \href{https://doi.org/10.1007/JHEP09(2022)201}{\emph{JHEP} {\bfseries 09}
  (2022) 201} [\href{https://arxiv.org/abs/2207.14224}{{\ttfamily
  2207.14224}}].

\bibitem{Heller:2022ejw}
M.~P. Heller, A.~Serantes, M.~Spali\'nski and B.~Withers, \emph{{Rigorous
  Bounds on Transport from Causality}},
  \href{https://doi.org/10.1103/PhysRevLett.130.261601}{\emph{Phys. Rev. Lett.}
  {\bfseries 130} (2023) 261601}
  [\href{https://arxiv.org/abs/2212.07434}{{\ttfamily 2212.07434}}].

\bibitem{Creminelli:2024lhd}
P.~Creminelli, O.~Janssen, B.~Salehian and L.~Senatore, \emph{{Positivity
  bounds on electromagnetic properties of media}},
  \href{https://doi.org/10.1007/JHEP08(2024)066}{\emph{JHEP} {\bfseries 08}
  (2024) 066} [\href{https://arxiv.org/abs/2405.09614}{{\ttfamily
  2405.09614}}].

\bibitem{Gavassino:2023mad}
L.~Gavassino, M.~M. Disconzi and J.~Noronha, \emph{{Dispersion Relations Alone
  Cannot Guarantee Causality}},
  \href{https://doi.org/10.1103/PhysRevLett.132.162301}{\emph{Phys. Rev. Lett.}
  {\bfseries 132} (2024) 162301}
  [\href{https://arxiv.org/abs/2307.05987}{{\ttfamily 2307.05987}}].

\bibitem{Stein}
E.~M. Stein and R.~Shakarchi, \emph{Complex Analysis}, Princeton Lectures in
  Analysis. Princeton University Press, 2003.

\bibitem{Bogolubov:1990ask}
N.~N. Bogolubov, A.~A. Logunov, A.~I. Oksak and I.~T. Todorov, eds.,
  \emph{{General Principles of Quantum Field Theory}}, vol.~10 of
  \emph{Mathematical Physics and Applied Mathematics}. Springer, 1990,
  \href{https://doi.org/10.1007/978-94-009-0491-0}{10.1007/978-94-009-0491-0}.

\bibitem{Kravchuk:2021kwe}
P.~Kravchuk, J.~Qiao and S.~Rychkov, \emph{{Distributions in CFT. Part II.
  Minkowski space}}, \href{https://doi.org/10.1007/JHEP08(2021)094}{\emph{JHEP}
  {\bfseries 08} (2021) 094}
  [\href{https://arxiv.org/abs/2104.02090}{{\ttfamily 2104.02090}}].

\bibitem{Witten:2023qsv}
E.~Witten, \emph{{Algebras, regions, and observers.}},
  \href{https://doi.org/10.1090/pspum/107/01954}{\emph{Proc. Symp. Pure Math.}
  {\bfseries 107} (2024) 247}
  [\href{https://arxiv.org/abs/2303.02837}{{\ttfamily 2303.02837}}].

\bibitem{Hollowood:2007kt}
T.~J. Hollowood and G.~M. Shore, \emph{{Causality and Micro-Causality in Curved
  Spacetime}},
  \href{https://doi.org/10.1016/j.physletb.2007.08.073}{\emph{Phys. Lett. B}
  {\bfseries 655} (2007) 67} [\href{https://arxiv.org/abs/0707.2302}{{\ttfamily
  0707.2302}}].

\bibitem{Hollowood:2007ku}
T.~J. Hollowood and G.~M. Shore, \emph{{The Refractive index of curved
  spacetime: The Fate of causality in QED}},
  \href{https://doi.org/10.1016/j.nuclphysb.2007.11.034}{\emph{Nucl. Phys. B}
  {\bfseries 795} (2008) 138}
  [\href{https://arxiv.org/abs/0707.2303}{{\ttfamily 0707.2303}}].

\bibitem{Hollowood:2015elj}
T.~J. Hollowood and G.~M. Shore, \emph{{Causality Violation, Gravitational
  Shockwaves and UV Completion}},
  \href{https://doi.org/10.1007/JHEP03(2016)129}{\emph{JHEP} {\bfseries 03}
  (2016) 129} [\href{https://arxiv.org/abs/1512.04952}{{\ttfamily
  1512.04952}}].

\bibitem{Manohar:2018aog}
A.~V. Manohar, \emph{{Introduction to Effective Field Theories}},
  \href{https://arxiv.org/abs/1804.05863}{{\ttfamily 1804.05863}}.

\bibitem{Adams:2006sv}
A.~Adams, N.~Arkani-Hamed, S.~Dubovsky, A.~Nicolis and R.~Rattazzi,
  \emph{{Causality, analyticity and an IR obstruction to UV completion}},
  \href{https://doi.org/10.1088/1126-6708/2006/10/014}{\emph{JHEP} {\bfseries
  10} (2006) 014} [\href{https://arxiv.org/abs/hep-th/0602178}{{\ttfamily
  hep-th/0602178}}].

\bibitem{Jackson:1998nia}
J.~D. Jackson, \emph{{Classical Electrodynamics}}. Wiley, 1998.

\bibitem{Landau9}
E.~M. Lifshitz and L.~P. Pitaevskii, \emph{Statistical physics: theory of the
  condensed state}, vol.~9. Elsevier, 2013.

\bibitem{Hui:2023pxc}
L.~Hui, I.~Kourkoulou, A.~Nicolis, A.~Podo and S.~Zhou, \emph{{S-matrix
  positivity without Lorentz invariance: a case study}},
  \href{https://doi.org/10.1007/JHEP04(2024)145}{\emph{JHEP} {\bfseries 04}
  (2024) 145} [\href{https://arxiv.org/abs/2312.08440}{{\ttfamily
  2312.08440}}].

\bibitem{Creminelli:2023kze}
P.~Creminelli, M.~Delladio, O.~Janssen, A.~Longo and L.~Senatore,
  \emph{{Non-analyticity of the S-matrix with spontaneously broken Lorentz
  invariance}}, \href{https://doi.org/10.1007/JHEP06(2024)201}{\emph{JHEP}
  {\bfseries 06} (2024) 201}
  [\href{https://arxiv.org/abs/2312.08441}{{\ttfamily 2312.08441}}].

\bibitem{Esposito:2020wsn}
A.~Esposito, R.~Krichevsky and A.~Nicolis, \emph{{Solidity without
  inhomogeneity: Perfectly homogeneous, weakly coupled, UV-complete solids}},
  \href{https://doi.org/10.1007/JHEP11(2020)021}{\emph{JHEP} {\bfseries 11}
  (2020) 021} [\href{https://arxiv.org/abs/2004.11386}{{\ttfamily
  2004.11386}}].

\bibitem{Nicolis:2015sra}
A.~Nicolis, R.~Penco, F.~Piazza and R.~Rattazzi, \emph{{Zoology of condensed
  matter: Framids, ordinary stuff, extra-ordinary stuff}},
  \href{https://doi.org/10.1007/JHEP06(2015)155}{\emph{JHEP} {\bfseries 06}
  (2015) 155} [\href{https://arxiv.org/abs/1501.03845}{{\ttfamily
  1501.03845}}].

\bibitem{Blas:2011ni}
D.~Blas and S.~Sibiryakov, \emph{{Horava gravity versus thermodynamics: The
  Black hole case}},
  \href{https://doi.org/10.1103/PhysRevD.84.124043}{\emph{Phys. Rev. D}
  {\bfseries 84} (2011) 124043}
  [\href{https://arxiv.org/abs/1110.2195}{{\ttfamily 1110.2195}}].

\bibitem{Creminelli:2012xb}
P.~Creminelli, J.~Norena, M.~Pena and M.~Simonovic, \emph{{Khronon inflation}},
  \href{https://doi.org/10.1088/1475-7516/2012/11/032}{\emph{JCAP} {\bfseries
  11} (2012) 032} [\href{https://arxiv.org/abs/1206.1083}{{\ttfamily
  1206.1083}}].

\bibitem{Kovtun:2012rj}
P.~Kovtun, \emph{{Lectures on hydrodynamic fluctuations in relativistic
  theories}}, \href{https://doi.org/10.1088/1751-8113/45/47/473001}{\emph{J.
  Phys. A} {\bfseries 45} (2012) 473001}
  [\href{https://arxiv.org/abs/1205.5040}{{\ttfamily 1205.5040}}].

\bibitem{Joyce:2022ydd}
A.~Joyce, A.~Nicolis, A.~Podo and L.~Santoni, \emph{{Integrating out beyond
  tree level and relativistic superfluids}},
  \href{https://doi.org/10.1007/JHEP09(2022)066}{\emph{JHEP} {\bfseries 09}
  (2022) 066} [\href{https://arxiv.org/abs/2204.03678}{{\ttfamily
  2204.03678}}].

\bibitem{Nicolis:2023pye}
A.~Nicolis, A.~Podo and L.~Santoni, \emph{{The connection between nonzero
  density and spontaneous symmetry breaking for interacting scalars}},
  \href{https://doi.org/10.1007/JHEP09(2023)200}{\emph{JHEP} {\bfseries 09}
  (2023) 200} [\href{https://arxiv.org/abs/2305.08896}{{\ttfamily
  2305.08896}}].

\bibitem{Baier:2008}
R.~Baier, P.~Romatschke, D.~T. Son, A.~O. Starinets and M.~A. Stephanov,
  \emph{Relativistic viscous hydrodynamics, conformal invariance, and
  holography},
  \href{https://doi.org/10.1088/1126-6708/2008/04/100}{\emph{Journal of High
  Energy Physics} {\bfseries 2008} (2008) 100}.

\bibitem{Landau6}
L.~D. Landau and E.~M. Lifshitz, \emph{Fluid Mechanics: Volume 6}, vol.~6.
  Elsevier, 1987.

\bibitem{Baumann:2019ghk}
D.~Baumann, D.~Green and T.~Hartman, \emph{{Dynamical Constraints on RG Flows
  and Cosmology}}, \href{https://doi.org/10.1007/JHEP12(2019)134}{\emph{JHEP}
  {\bfseries 12} (2019) 134}
  [\href{https://arxiv.org/abs/1906.10226}{{\ttfamily 1906.10226}}].

\bibitem{deRham:2021bll}
C.~de~Rham, A.~J. Tolley and J.~Zhang, \emph{{Causality Constraints on
  Gravitational Effective Field Theories}},
  \href{https://doi.org/10.1103/PhysRevLett.128.131102}{\emph{Phys. Rev. Lett.}
  {\bfseries 128} (2022) 131102}
  [\href{https://arxiv.org/abs/2112.05054}{{\ttfamily 2112.05054}}].

\bibitem{Green:2023ids}
D.~Green, Y.~Huang, C.-H. Shen and D.~Baumann, \emph{{Positivity from
  Cosmological Correlators}},
  \href{https://doi.org/10.1007/JHEP04(2024)034}{\emph{JHEP} {\bfseries 04}
  (2024) 034} [\href{https://arxiv.org/abs/2310.02490}{{\ttfamily
  2310.02490}}].

\bibitem{CarrilloGonzalez:2023emp}
M.~Carrillo~Gonz\'alez, \emph{{Bounds on EFT\textquoteright{}s in an expanding
  universe}}, \href{https://doi.org/10.1103/PhysRevD.109.085008}{\emph{Phys.
  Rev. D} {\bfseries 109} (2024) 085008}
  [\href{https://arxiv.org/abs/2312.07651}{{\ttfamily 2312.07651}}].

\bibitem{Donath:2024utn}
Y.~Donath and E.~Pajer, \emph{{The in-out formalism for in-in correlators}},
  \href{https://doi.org/10.1007/JHEP07(2024)064}{\emph{JHEP} {\bfseries 07}
  (2024) 064} [\href{https://arxiv.org/abs/2402.05999}{{\ttfamily
  2402.05999}}].

\bibitem{Cheung:2007st}
C.~Cheung, P.~Creminelli, A.~L. Fitzpatrick, J.~Kaplan and L.~Senatore,
  \emph{{The Effective Field Theory of Inflation}},
  \href{https://doi.org/10.1088/1126-6708/2008/03/014}{\emph{JHEP} {\bfseries
  03} (2008) 014} [\href{https://arxiv.org/abs/0709.0293}{{\ttfamily
  0709.0293}}].

\bibitem{Weinberg:2008hq}
S.~Weinberg, \emph{{Effective Field Theory for Inflation}},
  \href{https://doi.org/10.1103/PhysRevD.77.123541}{\emph{Phys. Rev. D}
  {\bfseries 77} (2008) 123541}
  [\href{https://arxiv.org/abs/0804.4291}{{\ttfamily 0804.4291}}].

\bibitem{Bordin:2017hal}
L.~Bordin, G.~Cabass, P.~Creminelli and F.~Vernizzi, \emph{{Simplifying the EFT
  of Inflation: generalized disformal transformations and redundant
  couplings}}, \href{https://doi.org/10.1088/1475-7516/2017/09/043}{\emph{JCAP}
  {\bfseries 09} (2017) 043}
  [\href{https://arxiv.org/abs/1706.03758}{{\ttfamily 1706.03758}}].

\bibitem{Visser:1998ua}
M.~Visser, B.~Bassett and S.~Liberati, \emph{{Superluminal censorship}},
  \href{https://doi.org/10.1016/S0920-5632(00)00782-9}{\emph{Nucl. Phys. B
  Proc. Suppl.} {\bfseries 88} (2000) 267}
  [\href{https://arxiv.org/abs/gr-qc/9810026}{{\ttfamily gr-qc/9810026}}].

\bibitem{Gao:2000ga}
S.~Gao and R.~M. Wald, \emph{{Theorems on gravitational time delay and related
  issues}}, \href{https://doi.org/10.1088/0264-9381/17/24/305}{\emph{Class.
  Quant. Grav.} {\bfseries 17} (2000) 4999}
  [\href{https://arxiv.org/abs/gr-qc/0007021}{{\ttfamily gr-qc/0007021}}].

\bibitem{Caron-Huot:2021rmr}
S.~Caron-Huot, D.~Mazac, L.~Rastelli and D.~Simmons-Duffin, \emph{{Sharp
  boundaries for the swampland}},
  \href{https://doi.org/10.1007/JHEP07(2021)110}{\emph{JHEP} {\bfseries 07}
  (2021) 110} [\href{https://arxiv.org/abs/2102.08951}{{\ttfamily
  2102.08951}}].

\bibitem{Birrell:1982ix}
N.~D. Birrell and P.~C.~W. Davies, \emph{{Quantum Fields in Curved Space}},
  Cambridge Monographs on Mathematical Physics. Cambridge University Press,
  Cambridge, UK, 1982,
  \href{https://doi.org/10.1017/CBO9780511622632}{10.1017/CBO9780511622632}.

\bibitem{Weinberg:1995mt}
S.~Weinberg, \emph{{The Quantum theory of fields. Vol. 1: Foundations}}.
  Cambridge University Press, 6, 2005,
  \href{https://doi.org/10.1017/CBO9781139644167}{10.1017/CBO9781139644167}.

\end{thebibliography}\endgroup

\end{document}